\documentclass[aps,prb,twocolumn,superscriptaddress]{revtex4-2}

\usepackage{amsmath,amssymb}
\usepackage{graphicx}
\usepackage{dcolumn}
\usepackage{bm}
\usepackage{siunitx}
\usepackage{booktabs}
\usepackage{longtable}
\usepackage{xcolor}
\usepackage{placeins}
\usepackage{microtype}
\usepackage{hyperref}
\usepackage{capt-of}
\usepackage{float}
\usepackage[version=4]{mhchem}
\usepackage{makecell}
\usepackage{multirow}

\hypersetup{
  colorlinks=true,
  linkcolor=blue,
  citecolor=blue,
  urlcolor=blue
}

\begin{document}

\title{Revealing the Role of Confined Molecular \ce{H2} in the Passivation of Defective Silicon Using First-Principles Simulations}

\author{Hania Azzam}
%\email[Corresponding author: ]{h.azzam@fz-juelich.de}
\affiliation{Theory and Computation of Energy Materials (IET-3), Forschungszentrum J\"ulich, 52425 J\"ulich, Germany}
\affiliation{Chair of Theory and Computation of Energy Materials, Faculty of Georesources and Materials Engineering, RWTH Aachen University, 52062 Aachen, Germany}

\author{Tobias Binninger}
\email[Corresponding author: ]{t.binninger@fz-juelich.de}
\affiliation{Theory and Computation of Energy Materials (IET-3), Forschungszentrum J\"ulich, 52425 J\"ulich, Germany}

\author{Benedikt Fischer}
\author{Uwe Rau}
\affiliation{Energy Materials and Devices - Photovoltaics (IMD-3), Forschungszentrum J\"ulich, 52425 J\"ulich, Germany}

\author{Michael Eikerling}
\affiliation{Theory and Computation of Energy Materials (IET-3), Forschungszentrum J\"ulich, 52425 J\"ulich, Germany}
\affiliation{Chair of Theory and Computation of Energy Materials, Faculty of Georesources and Materials Engineering, RWTH Aachen University, 52062 Aachen, Germany}

\begin{abstract}
The passivation of silicon dangling bonds by hydrogen is a crucial requirement for silicon-based optoelectronic technology, especially for solar cells. Recent experiments on intense light soaking of silicon heterojunction solar cells unveiled interesting dynamical aspects of hydrogen passivation that are linked to Si-H bond breaking and repassivation. These processes take place predominantly in porous regions near the amorphous/crystalline interface, where hydrogen can exist in molecular form. This work addresses the question of whether molecular \(\ce{H2}\) directly participates in Si--H depassivation and repassivation. Using density functional theory, we calculate and compare formation energies of point defects, multivacancy cavities and the Si(100) surface to identify relevant passivated and depassivated states. Furthermore, we employ nudged elastic band calculations to determine the activation barriers of the corresponding pathways. We find that sufficient local free volume enables a direct double-H pathway for depassivation through the formation of confined molecular \(\ce{H2}\). Despite involving the breaking of two Si--H bonds, the double-H process can be energetically and kinetically competitive with the single-H process and can exhibit a reverse repassivation barrier as low as \(0.15\,\mathrm{eV}\) under p-type conditions. These findings provide a plausible atomistic explanation for passivation recovery during light soaking and illuminated annealing in porous regions near amorphous/crystalline silicon interfaces.
\end{abstract}

\maketitle

% ============================================================
\section{Introduction}
\label{sec:introduction}
% ============================================================

Hydrogen plays an important role in silicon-based photovoltaics by passivating electrically active defects through the formation of Si--H bonds \cite{VanDeWalle1994DB,VanDeWalle1998,Hallam2020,Hammann2025}. Unpassivated dangling Si bonds act as recombination centers for photogenerated charge carriers and are thus detrimental for the photovoltaic performance \cite{VanDeWalle1994DB,Hallam2020,Chen2021a,Hammann2025}. The formation of dangling Si bonds via Si--H bond breaking contributes to long-term degradation of silicon-based photovoltaic devices \cite{VanDeWalle1994DB,Chen2021a,Diggs2023HydrogenDegradation,Diggs2025FermiLevel}. On the other hand, trapping of hydrogen in crystalline or amorphous silicon can help to regenerate the passivated state of Si--H bonds, a process that was suggested to contribute to the experimentally established performance enhancement under so-called light-soaking conditions, i.e., the simultaneous exposure of the material to elevated temperatures and strong illumination \cite{Mahtani2013,Kobayashi2016APL,Hallam2020,Chen2021a,Hammann2023JPHOTOV,Fischer2023,Diggs2023HydrogenDegradation,Duan2024SolarRRL,Hammann2024SolarRRL,Hammann2025,Fischer2025CellReports,Lajoie2025,Kwapil2026LightSoaking}. The competition between loss and regeneration of defect passivation is thus directly related to device performance and durability \cite{Hallam2020,Chen2021a,Fischer2023,Hammann2025,Lajoie2025}. A microscopic understanding of passivation stability in silicon not only requires identifying the lowest-energy hydrogenated structures, but also clarifying how Si--H passivation can be lost or recovered under different local environments \cite{VanDeWalle1994,VanDeWalle1998,VanDeWalleNeugebauer2006}.

Recent work on underdense a-Si:H and silicon heterojunction (SHJ) solar-cell degradation has sharpened the broader device-level context. Fischer \emph{et al.} showed that the microstructure of underdense hydrogenated amorphous silicon is directly relevant to silicon heterojunction passivation and argued that molecular hydrogen diffusion through a void network can play an important role in passivation for underdense a-Si:H \cite{Fischer2023}. Light soaking and carrier injection studies have also shown that illumination or forward bias can improve passivation and device performance in SHJ cells \cite{Kobayashi2016APL,Duan2024SolarRRL,Fischer2025CellReports}. Other studies on UV-induced degradation have highlighted the roles of hydrogen redistribution and doping in passivation loss and reversibility under nonequilibrium conditions \cite{Lajoie2025}. These studies do not resolve the atomistic reaction pathways addressed here, but they underscore that hydrogen chemistry in imperfect Si environments is central to passivation stability.

Prior first-principles computational work has clarified the basic hydrogen-related structures and energetics in several structural environments. In bulk crystalline silicon, atomic interstitial hydrogen and molecular (di-)hydrogen have both been examined in detail, and molecular \(\ce{H2}\) was shown to represent an important low-energy hydrogen configuration \cite{VanDeWalle1994,Herring2001,VanDeWalle1998,VanDeWalleNeugebauer2006}. Point silicon defects such as the Si monovacancy and divacancy have likewise been studied, including the sequential hydrogenation of vacancy structures and the stabilization of the resulting hydrogen-passivated structures \cite{VanDeWalle1994DB,Coutinho2003,Ghasemi2014,Kolevatov2019}. At larger Si vacancy clusters, Akiyama and Oshiyama showed that multivacancies, such as six-fold \(\mathrm{V_6}\) and ten-fold \(\mathrm{V_{10}}\) Si-atom vacancies, can trap and store molecular \(\ce{H2}\) within the void space \cite{AkiyamaOshiyama2001,Hourahine1999,Ishioka1999,Mori2001}. These works established vacancy cavities as stable sites for molecular hydrogen. However, they did not address partially depassivated cavities in which Si--H bonds are broken, dangling bonds are formed, and the released hydrogen forms confined \(\ce{H2}\).
External Si surfaces, especially Si(100), have also served as standard systems for the computational study of Si--H bond breaking and associative hydrogen desorption \cite{Hoefer1992,Kratzer1995,Lin1999,Durr2006,Tsatsoulis2018a,Zhou2022}. Together, these studies establish the main hydrogen configurations and stabilities in crystalline silicon, point defects, multivacancies, and Si surfaces. Transition-state calculations by Diggs \emph{et al.} provide insights into the kinetics of the depassivation process, where one Si--H bond breaks and atomic hydrogen moves away from the defect \cite{DiggsThesis2025,Diggs2023HydrogenDegradation,Diggs2025FermiLevel}. What remains unclear is whether molecular \(\ce{H2}\) can also take part directly in local Si--H depassivation and repassivation. In particular, can two H atoms released by breaking neighboring Si--H bonds form a local \(\ce{H2}\) molecule, and can this molecule later repassivate the resulting dangling bonds? This question is important because \(\ce{H2}\) is often discussed as a stored hydrogen species, whereas its possible direct participation in degradation and recovery has received little attention. 
To address this question, we compare the single-H depassivation process with a double-H process. Here, the double-H process refers to breaking two neighboring Si--H bonds and forming one \(\ce{H2}\) molecule from the released H atoms. We study these processes in a point defect, in multivacancy cavities, and at the Si(100) surface.

The remainder of this paper is organized as follows. Section~\ref{sec:methods} describes the computational approach and thermodynamic framework. Section~\ref{sec:configurations} defines the configuration space and the corresponding passivated and depassivated structures. Section~\ref{sec:results} presents the energetic stability and reaction barriers of single-H and double-H processes. Section~\ref{sec:discussion} discusses the implications of the computational results for understanding the microscopic mechanisms underlying hydrogen stability and recovery in defective silicon. These findings are further discussed in the context of photovoltaic passivation and related technologies.

% ============================================================
\section{Computational Methods}
\label{sec:methods}
% ============================================================

\subsection{Density Functional Theory}
\label{sec:dft}

Electronic structure calculations based on density functional theory (DFT) were performed using the \textsc{Quantum ESPRESSO} package \cite{Giannozzi2009QE,Giannozzi2017QE}. Exchange and correlation were treated within the generalized-gradient approximation using the Perdew--Burke--Ernzerhof (PBE) functional \cite{Perdew1996PBE}. Interactions with core-level electrons were described using projector augmented wave (PAW) pseudopotentials \cite{Kresse1999PAW} from the \texttt{pslibrary 1.0.0} set for both Si and H \cite{DalCorso2014PSLibrary}. Wave functions were expanded in a plane-wave basis with a kinetic energy cutoff of \(50\,\mathrm{Ry}\) and a charge density cutoff of \(400\,\mathrm{Ry}\). For pristine and defective bulk silicon, we employed a \(2\times2\times2\) supercell of the conventional 8-atom unit cell, containing 64 Si atoms in the pristine reference structure. The relaxed lattice constant of the cubic unit cell was \(a=5.469\,\mathrm{\AA}\), giving a supercell length of \(10.938\,\mathrm{\AA}\). The Brillouin-zone integration was performed with a Monkhorst--Pack \(4\times4\times4\) \(k\)-point mesh \cite{MonkhorstPack1976}. Kohn-Sham orbital occupations were treated using Fermi--Dirac smearing with \(\sigma=0.001\,\mathrm{Ry}\). These settings were established from convergence tests on total energies and forces, and benchmarked against the experimental lattice constant of crystalline Si, \(a=5.431\,\mathrm{\AA}\) \cite{OkadaTokumaru1984}.

All calculations were carried out spin-polarized to capture partially passivated configurations with dangling bonds. While pristine silicon and fully hydrogenated structures relax to nonmagnetic states, structures with one or more dangling bonds may exhibit open-shell character and require spin-polarized treatment to obtain reliable total energies. A small initial magnetization was used to seed symmetry breaking around defect sites. The electronic self-consistency threshold was set to \(10^{-7}\,\mathrm{Ry}\). Structural relaxations were performed using the BFGS algorithm until the change of the total energy was below \(10^{-5}\,\mathrm{Ry}\) and the forces were below \(10^{-4}\,\mathrm{Ry/bohr}\).

For the relaxed configurations, the electronic states were further characterized using projected densities of states (PDOS) calculated with the \texttt{projwfc.x} module of Quantum ESPRESSO. The PDOS were used to resolve the contributions of individual atoms to the electronic states near the Fermi level, particularly those associated with the created Si dangling bonds and the bond-centered or molecular H configurations. Löwdin atomic populations were obtained by projecting the Kohn--Sham wave functions onto orthogonalized atomic wave functions \cite{lowdin1950} and were used to examine the redistribution of electronic population among the atoms involved in the defect configurations.

For the surface calculations, we employed a symmetric 6-layer Si(100) slab containing 48 atoms in a \(15.470 \times 7.735\,{\AA}^2\) in-plane supercell. The simulation cell length normal to the surface is \(28.20\,\AA\), providing a vacuum spacing of \(21\,\AA\) sufficient to suppress slab--slab interactions. The bulk terminated Si(100) slab was first relaxed without symmetry constraints, allowing neighboring surface Si atoms to form the characteristic buckled-dimer reconstruction of Si(100). This reconstruction lowers the surface energy by pairing two neighboring surface atoms, thereby reducing the number of dangling bonds from two to one per surface Si atom relative to the unreconstructed Si(100) surface. Buckling further stabilizes the reconstructed surface by allowing the two atoms in each dimer to relax to inequivalent vertical positions \cite{Ramstad1995Si100}. The remaining dangling bonds on the reconstructed surface were then saturated with H atoms to construct monohydride passivated Si(100) surface. The in-plane Brillouin zone was sampled with a \(4\times8\times1\) Monkhorst--Pack mesh \cite{MonkhorstPack1976}. During structural relaxation, the total energy convergence threshold was \(10^{-5}\,\mathrm{Ry}\) and the force convergence threshold was \(10^{-4}\,\mathrm{Ry/bohr}\).

\subsection{Thermodynamics: Grand-Canonical Framework}
\label{sec:chemical_potential}

To consistently compare hydrogen configurations across bulk, vacancy, cavity, and surface environments, formation energies were evaluated within a grand-canonical framework \cite{Freysoldt2014,Alkauskas2016}. Referencing to fixed chemical potentials virtually connects the simulation cell to reservoirs with which it can exchange atoms. Silicon is referenced to bulk crystalline silicon through
\begin{equation}
\mu_{\mathrm{Si}}=\frac{E_{\mathrm{tot}}(\mathrm{Si}_{64})}{64},
\label{eq:musi}
\end{equation}
where \(E_{\mathrm{tot}}(\mathrm{Si}_{64})\) is the DFT total energy of the pristine 64-atom Si supercell. Hydrogen is referenced to an isolated hydrogen molecule in the gas phase. The corresponding reference chemical potential is
\begin{equation}
\mu_{\mathrm{H}}^{0}=\frac{1}{2}E_{\mathrm{tot}}(\mathrm{H}_{2}),
\label{eq:muh0}
\end{equation}
with one H atom assigned one half of the DFT total energy of an isolated \(\ce{H2}\) molecule. The hydrogen chemical potential is then treated as a tunable parameter and written as 

\begin{equation}
\mu_{\mathrm H} = \mu_{\mathrm H}^{0} + \Delta\mu_{\mathrm H},
\label{eq:muh_delta}
\end{equation}
where \(\mu_{\mathrm H}\) describes how favorable it is for a defect structure to exchange hydrogen with the hydrogen reservoir. \(\Delta\mu_{\mathrm H}\) specifies this chemical potential relative to \(\mu_{\mathrm H}^{0}\). Varying \(\Delta\mu_{\mathrm H}\) changes the contribution of hydrogen to the formation energy, and therefore shows how the relative stability of configurations with different hydrogen content varies with the hydrogen chemical potential. 

All chemical potentials and formation energies reported here are based on DFT total energies. Finite-temperature, zero-point energy, vibrational, or entropy corrections have not been included.

The formation energy \(E_{\mathrm{f}}\) of a configuration \(X\) containing \(N_{\mathrm{Si}}\) silicon atoms and \(N_{\mathrm{H}}\) hydrogen atoms is written as
\begin{equation}
E_{\mathrm{f}}[X;\mu_{\mathrm{H}}]
=
E_{\mathrm{tot}}[X]
-
N_{\mathrm{Si}}\mu_{\mathrm{Si}}
-
N_{\mathrm{H}}\mu_{\mathrm{H}}.
\label{eq:eform_general}
\end{equation}

Equation~\eqref{eq:eform_general} is used throughout this work to rank candidate Si--H structures and identify the lowest energy states for each defect topology. For reactions in which the initial and final states contain the same number of silicon and hydrogen atoms, such as direct depassivation steps, the chemical-potential terms cancel and the grand-canonical energy difference reduces to the total energy difference between final and initial states,
\begin{equation}
\Delta E = E_{\mathrm{tot}}^{\mathrm{final}} - E_{\mathrm{tot}}^{\mathrm{initial}}.
\label{eq:reaction_energy}
\end{equation}

\subsection{Carrier-Type Modeling}
\label{sec:doping}

To study the influence of doping on the depassivation and repassivation pathways, we performed calculations for undoped, n-type and p-type cases. The n-type case was modeled by adding one electron to the supercell, while the p-type case was modeled by removing one electron. In periodic DFT calculations, a supercell with a net charge would lead to a divergent electrostatic energy because the charged cell is repeated periodically. This divergence is avoided by adding a homogeneous compensating background charge, often referred to as a jellium background, which is uniformly distributed over the simulation cell \cite{MakovPayne1995}. This uniform background charge can be interpreted as a simple delocalized model of the core atomic charge of some dopant heteroatoms.

The n-type and p-type calculations performed here should therefore not be interpreted as explicit dopant states. Instead, they are used to examine how an added electron or hole affects the reaction energies and activation barriers.

\subsection{Reaction Barriers and Kinetics}
\label{sec:neb_methods}

Minimum-energy pathways for selected depassivation and repassivation reactions were computed using the nudged elastic band (NEB) method \cite{Henkelman2000NEB,Henkelman2000ImprovedNEB} as implemented in \textsc{Quantum ESPRESSO}. Initial and final states were taken from the relaxed configurations defined in Sec.~\ref{sec:configurations} and identified from their formation energies, calculated as described in Sec.~\ref{sec:chemical_potential}. NEB calculations were used to determine both forward and reverse activation barriers for representative single-H and double-H (\ce{H2}) processes, including monovacancy depassivation by single Si--H bond breaking and the direct formation of trapped \(\ce{H2}\) by double Si--H bond breaking from nanocavity walls. The calculated barriers enable direct comparison of the kinetics of different depassivation and repassivation mechanisms across various defect topologies.

For each NEB pathway, a set of seven images was generated by linear interpolation between the fully relaxed initial and final states. A climbing-image scheme was employed to promote the highest-energy image to the transition state of the barrier. The path was optimized until the forces acting on the atoms in the intermediate NEB images were below \(0.05\,\mathrm{eV/\AA}\), indicating that the minimum-energy pathway was converged. Forward and reverse activation barriers were then obtained from the energy difference between the transition-state image and the initial and final states, respectively.

\subsection{Validation with Hybrid Functionals}
\label{sec:hse_methods}

Given the self-interaction errors and spurious delocalization associated with semilocal functionals such as PBE, hybrid-functional calculations with HSE06 \cite{Heyd2003,Heyd2006} were carried out as a check of the most relevant barrier estimates. Single-point HSE06 calculations were performed for the PBE-relaxed passivated and depassivated initial and final states of the \(\mathrm{V_1}\), \(\mathrm{V_6}\), as well as for the corresponding PBE-level transition-state geometries. These calculations were used to verify that the main conclusions of this work are robust against PBE artifacts. 

% ============================================================
\section{Configurations and Processes}
\label{sec:configurations}
% ============================================================

\subsection{Si Bulk Interstitial and Vacancy Structures}
\label{sec:bulk_vacancy_configurations}

\begin{figure}[htbp]
\centering
\includegraphics[width=0.45\columnwidth]{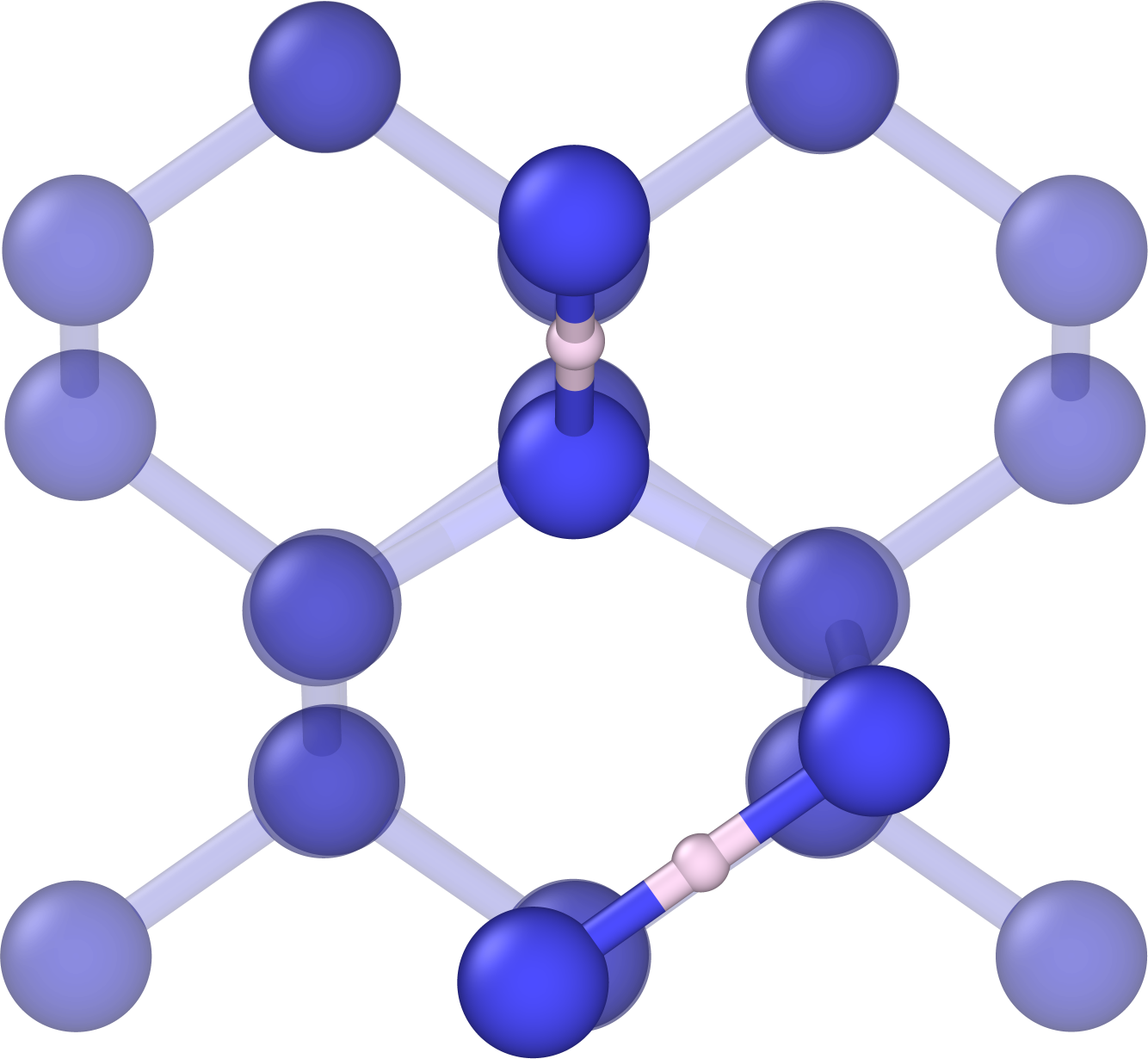}
\hspace{0.03\columnwidth}
\includegraphics[width=0.45\columnwidth]{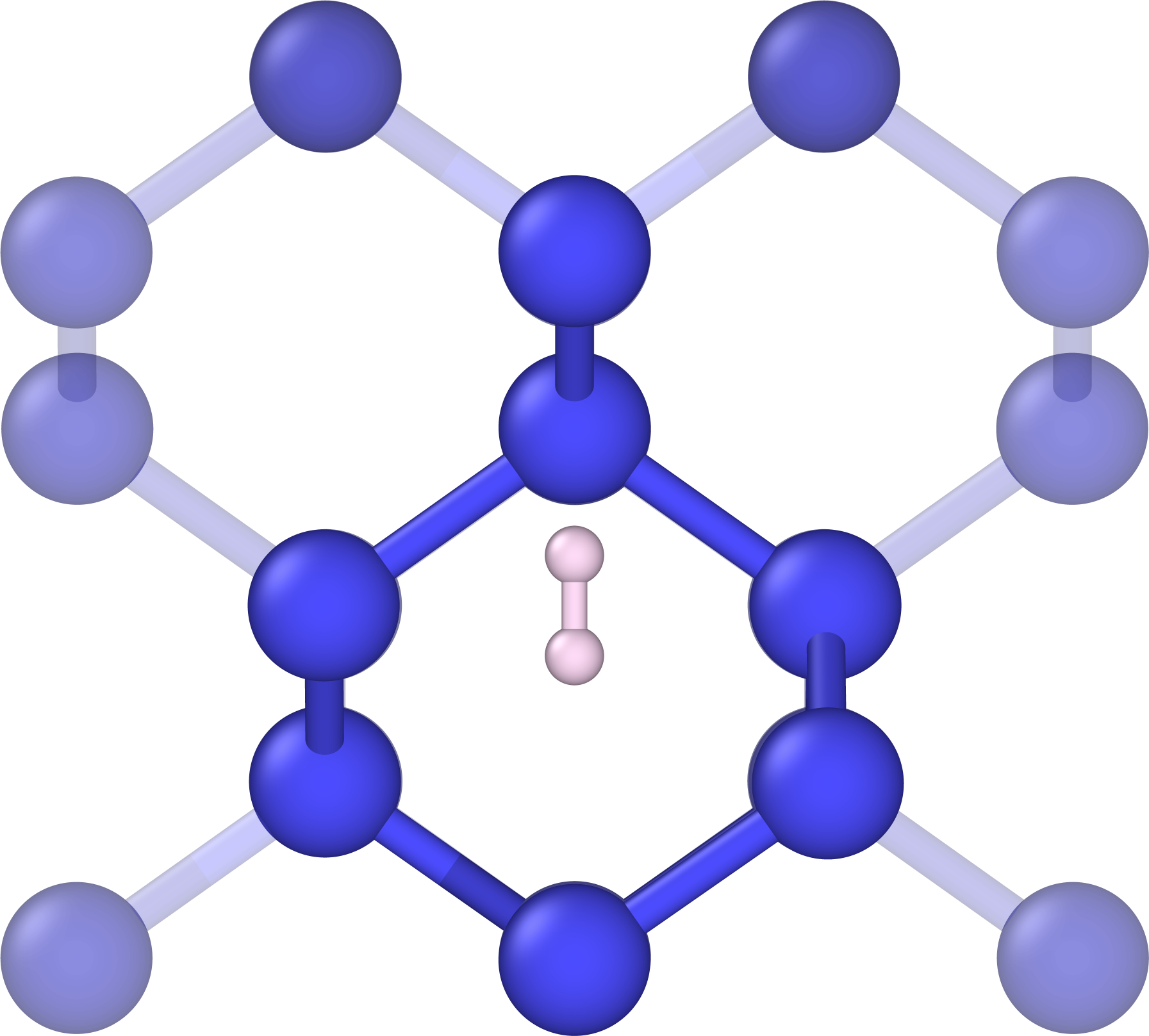}
\caption{Formation of molecular \(\ce{H2}\) in bulk Si. The initial state contains two bond-centered H atoms in crystalline Si, while the final state contains one molecular \(\ce{H2}\) molecule at a tetrahedral (T$_d$) interstitial site.}
\label{fig:bulk_structures}
\end{figure}

The bulk reference system in this work was crystalline silicon represented by a 64-atom supercell (\(\mathrm{Si}_{64}\)) from which both interstitial-hydrogen configurations and Si-vacancy--hydrogen defect structures were constructed. Within defect-free bulk silicon, three main sites for atomic interstitial hydrogen were considered: (i) the bond-centered (BC) site, located midway between a nearest-neighbor Si--Si bond and forming a three-center Si--H--Si bridge; (ii) the antibonding (AB) site, displaced along the Si--Si bond axis toward one Si atom; and (iii) the tetrahedral (T$_d$) interstitial site at the center of a tetrahedron formed by four Si atoms. These sites are commonly considered the most relevant configurations for interstitial hydrogen in crystalline Si \cite{VanDeWalle1994,Herring2001,VanDeWalle1998}. All atomic-H structures were fully relaxed. The bond-centered site, referred to as H(BC), was lowest in energy, consistent with previous first-principles studies of interstitial H in crystalline Si \cite{Herring2001}. Therefore, it is used as the atomic-H final state for the single-H depassivation reaction. In addition, we included molecular \(\ce{H2}\) states at a tetrahedral interstitial site (\(\ce{H2}(T_d)\)). The transformation of two atomic BC hydrogen to molecular \(\ce{H2}(T_d)\) is shown in Fig.~\ref{fig:bulk_structures} and written as
\begin{equation}
2\,\mathrm{H(BC)}
\rightarrow
\ce{H2}(T_d).
\label{eq:bulk_h2_reaction}
\end{equation}
This reaction was previously found to be energetically favorable in first-principles calculations, with interstitial \(\ce{H2}\) at a tetrahedral site lower in energy than two separated bond-centered H atoms in crystalline Si \cite{VanDeWalle1994}.

\begin{figure}[htbp]
\centering
\includegraphics[width=0.45\columnwidth]{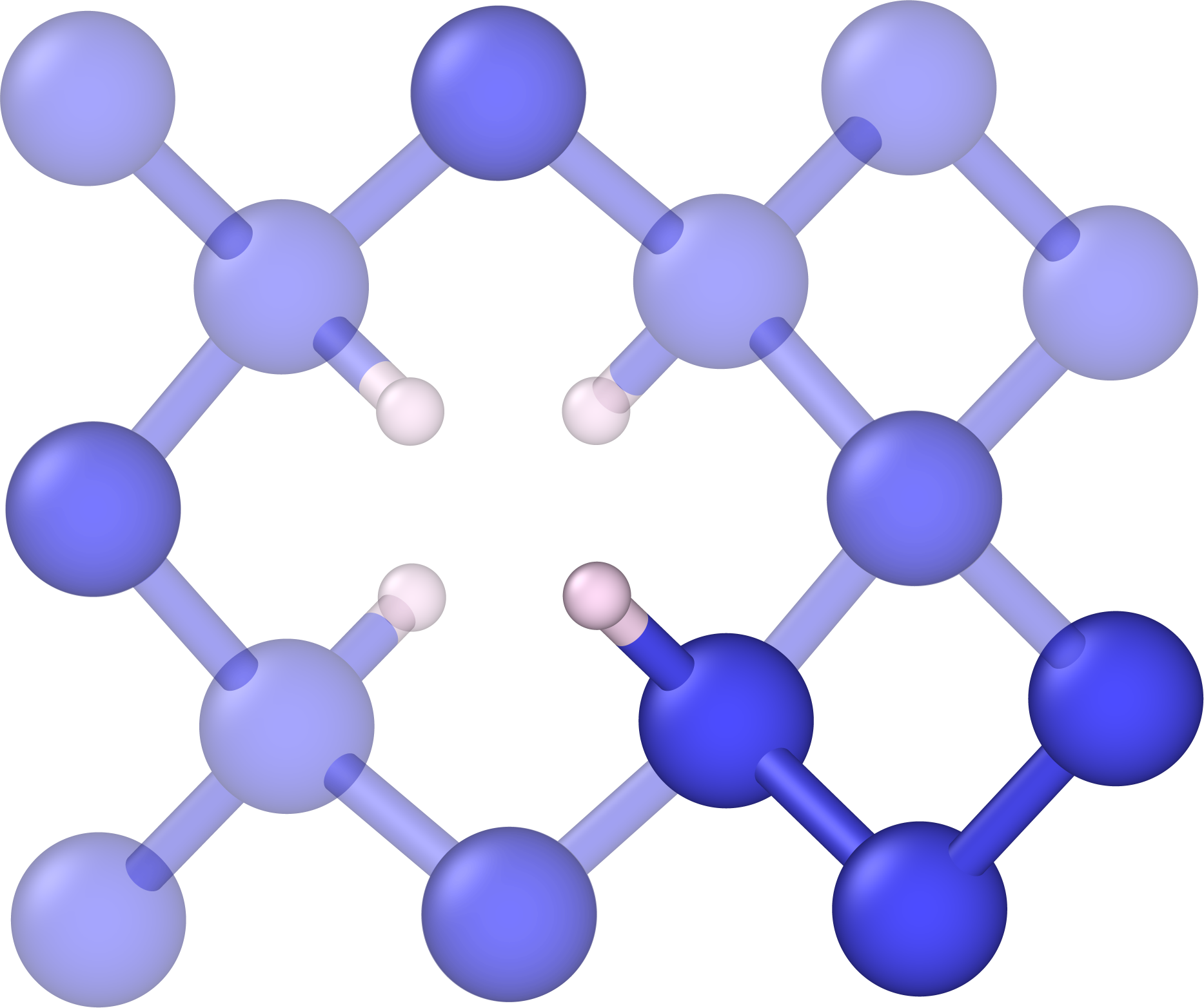}
\hspace{0.03\columnwidth}
\includegraphics[width=0.45\columnwidth]{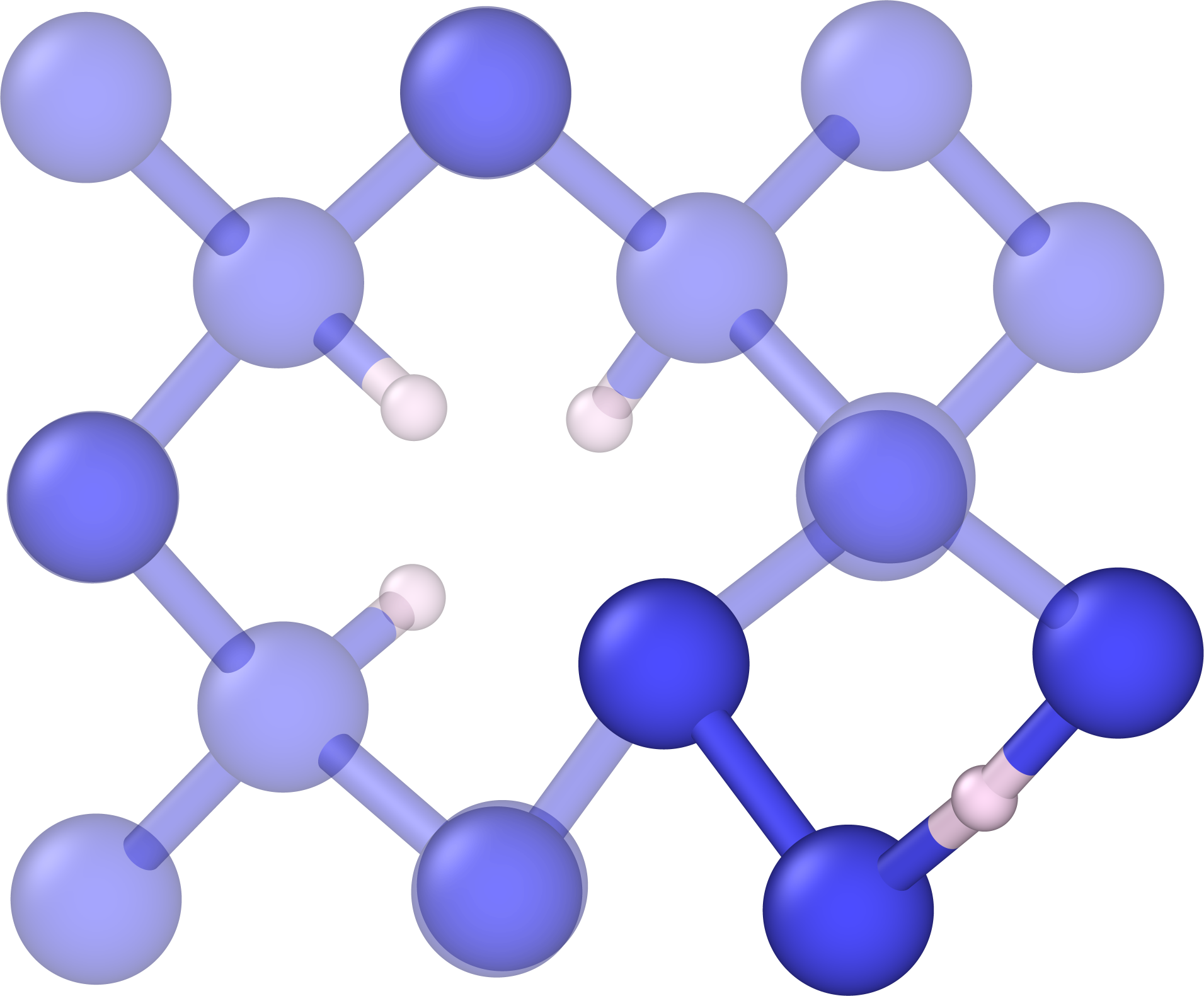}
\caption{Direct monovacancy depassivation. Initial structure (left) and final structure (right). The reacting H atom and the Si atom that forms a dangling bond in the final state are rendered fully opaque, while the remaining atoms are shown with reduced opacity for visual clarity.}
\label{fig:v1_structures}
\end{figure}

To model Si-vacancy--hydrogen defect structures, we introduce a single Si vacancy \(\mathrm{V_1}\) in the \(\mathrm{Si}_{64}\) supercell, thereby creating four dangling bonds (DBs) on the neighboring Si atoms. Sequential hydrogenation produces \(\mathrm{V_1H_n}\) configurations \((n=0\text{--}4)\), ranging from the bare vacancy to the fully saturated \(\mathrm{V_1H_4}\) structure, shown in Fig.~\ref{fig:v1_structures} (left). We further considered the direct depassivation step
\begin{equation}
\mathrm{V_1H_4}
\rightarrow
\mathrm{V_1H_3}+\mathrm{H(BC)}\ .
\label{eq:v1_depassivation}
\end{equation}
The same process can be written locally as 
\begin{equation}
\mathrm{Si\!-\!H}
\rightarrow
\mathrm{DB}+\mathrm{H(BC)},
\label{eq:single_sih_to_hbc}
\end{equation}
in which one Si--H bond is broken, creating one dangling Si bond, and the released hydrogen atom is placed in a BC interstitial site, as shown in Fig.~\ref{fig:v1_structures} (right). Diggs and co-workers recently studied related carrier-dependent single-H Si--H bond breaking processes with transition-state calculations \cite{DiggsThesis2025,Diggs2023HydrogenDegradation,Diggs2025FermiLevel}. This point-defect single-H pathway is then used for comparison with the double-H cavity pathway.

\subsection{Multivacancy and Cavity Configurations}
\label{sec:multivacancy_configurations}

To extend the analysis from single Si-vacancy point defects to nano-cavities, we constructed connected multivacancy clusters \(\mathrm{V}_n\) by removing \(n\) neighboring silicon atoms from the bulk \(\mathrm{Si}_{64}\) supercell. The subscript \(n\) denotes the number of removed Si atoms. The vacancy cluster geometries were chosen to remain compact and connected, so that the removed atoms form one local defect region rather than a set of isolated vacancies. This generated a series of defect structures ranging from point-like vacancies (\(\mathrm{V_1}\), \(\mathrm{V_2}\)) to larger internal cavities (\(\mathrm{V_4}\), \(\mathrm{V_6}\), and \(\mathrm{V_{10}}\)). These larger clusters provide enough internal void space to stabilize confined molecular \(\ce{H2}\), consistent with the first-principles study of hydrogen incorporation in Si multivacancies by Akiyama and Oshiyama \cite{AkiyamaOshiyama2001}. 

\begin{figure}[htbp]
\centering
\includegraphics[width=0.45\columnwidth]{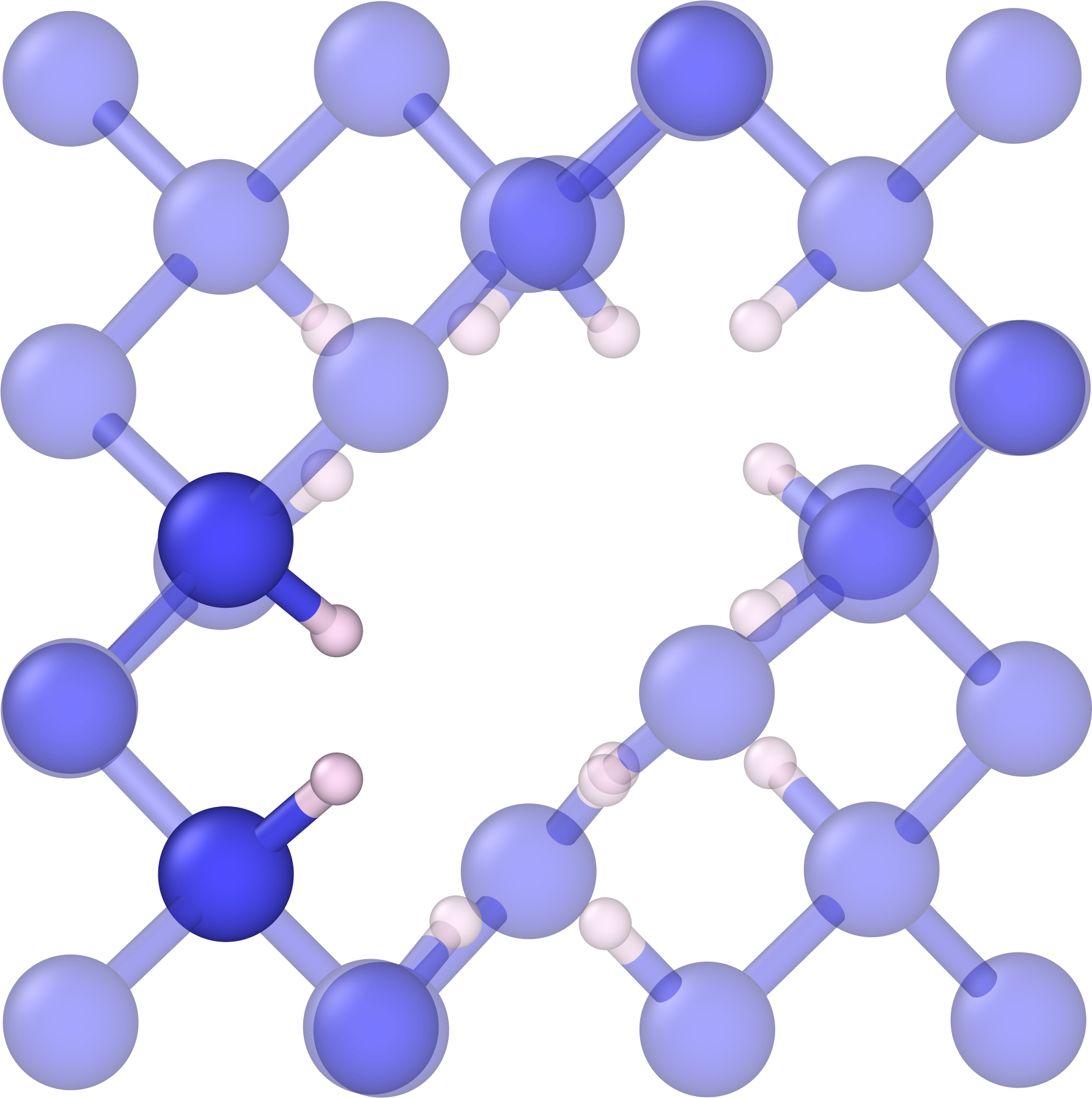}
\hspace{0.03\columnwidth}
\includegraphics[width=0.45\columnwidth]{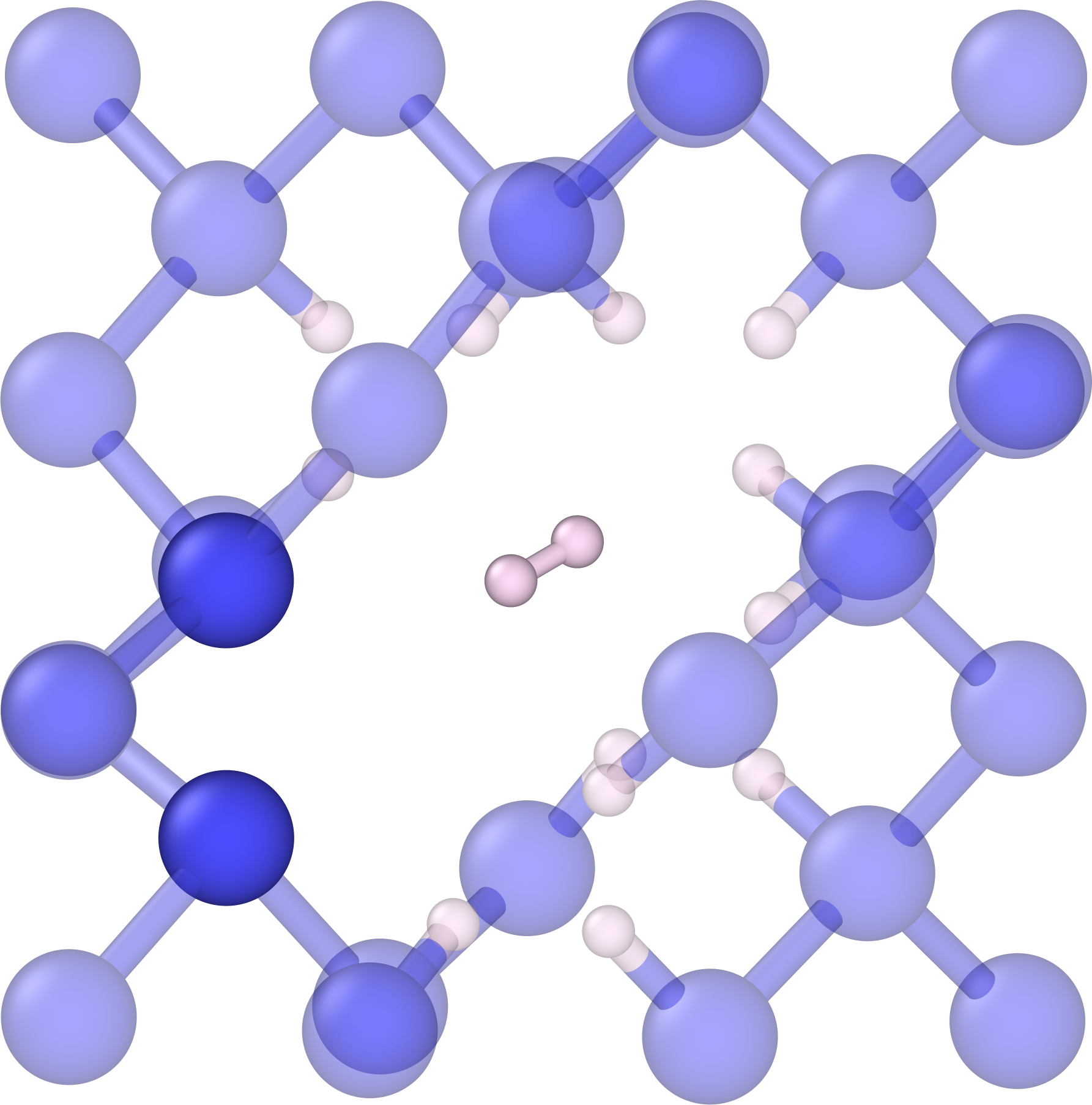}
\caption{\(\mathrm{V_6}\) cavity depassivation via direct \ce{H2} formation: the initial state is the fully passivated \(\mathrm{V_6H_{14}}\) cavity (left), while the final state contains two dangling bonds on the internal wall and one trapped \(\ce{H2}\) molecule inside the cavity (right). The reacting H atoms and the Si atoms that form dangling bonds in the final state are rendered fully opaque, while the remaining atoms are shown with reduced opacity for visual clarity.}
\label{fig:v6_structures}
\end{figure}

For each multivacancy cluster, we first constructed the fully hydrogen-passivated reference state by saturating all dangling bonds  on the cavity wall with hydrogen atoms. This gives \(\mathrm{V_4H_{10}}\), \(\mathrm{V_6H_{14}}\), and \(\mathrm{V_{10}H_{22}}\). We then constructed direct depassivation final states by breaking two Si--H bonds on the cavity wall and allowing the released hydrogen atoms to recombine into a single \(\ce{H2}\) molecule confined inside the cavity. For the \(\mathrm{V_6}\) cavity, the fully passivated initial state and the double-H depassivated state containing confined \(\mathrm{H_2}\) are shown in Fig.~\ref{fig:v6_structures}. The energy differences between these initial and final states were evaluated to quantify the cost of concerted cavity-wall depassivation and formation of confined molecular hydrogen. The respective process is
\begin{equation}
\mathrm{V_nH_m}
\rightarrow
\mathrm{V_nH_{m-2}}+\ce{H2_{conf}},
\label{eq:vn_double_h}
\end{equation}
where the resulting \(\ce{H2}\) remains confined inside the cavity. For the cavities considered here, this corresponds to
\(\mathrm{V_4H_{10}}\rightarrow\mathrm{V_4H_8}+\ce{H2_{conf}}\),
\(\mathrm{V_6H_{14}}\rightarrow\mathrm{V_6H_{12}}+\ce{H2_{conf}}\), and
\(\mathrm{V_{10}H_{22}}\rightarrow\mathrm{V_{10}H_{20}}+\ce{H2_{conf}}\).

The same reaction can also be written from the local bond breaking perspective as
\begin{equation}
2\,\mathrm{Si\!-\!H}
\rightarrow
2\,\mathrm{DB}+\ce{H2}.
\label{eq:two_sih_to_h2}
\end{equation}
This bond-level notation emphasizes that two Si--H bonds are converted into two Si dangling bonds and one H--H bond.

\subsection{Extended Surface Limit: Si(100)}
\label{sec:surface_configurations}

\begin{figure}[htbp]
\centering
\vspace{3em}
\begin{minipage}[c][0.34\columnwidth][c]{0.45\columnwidth}
\centering
\includegraphics[width=\linewidth,keepaspectratio]{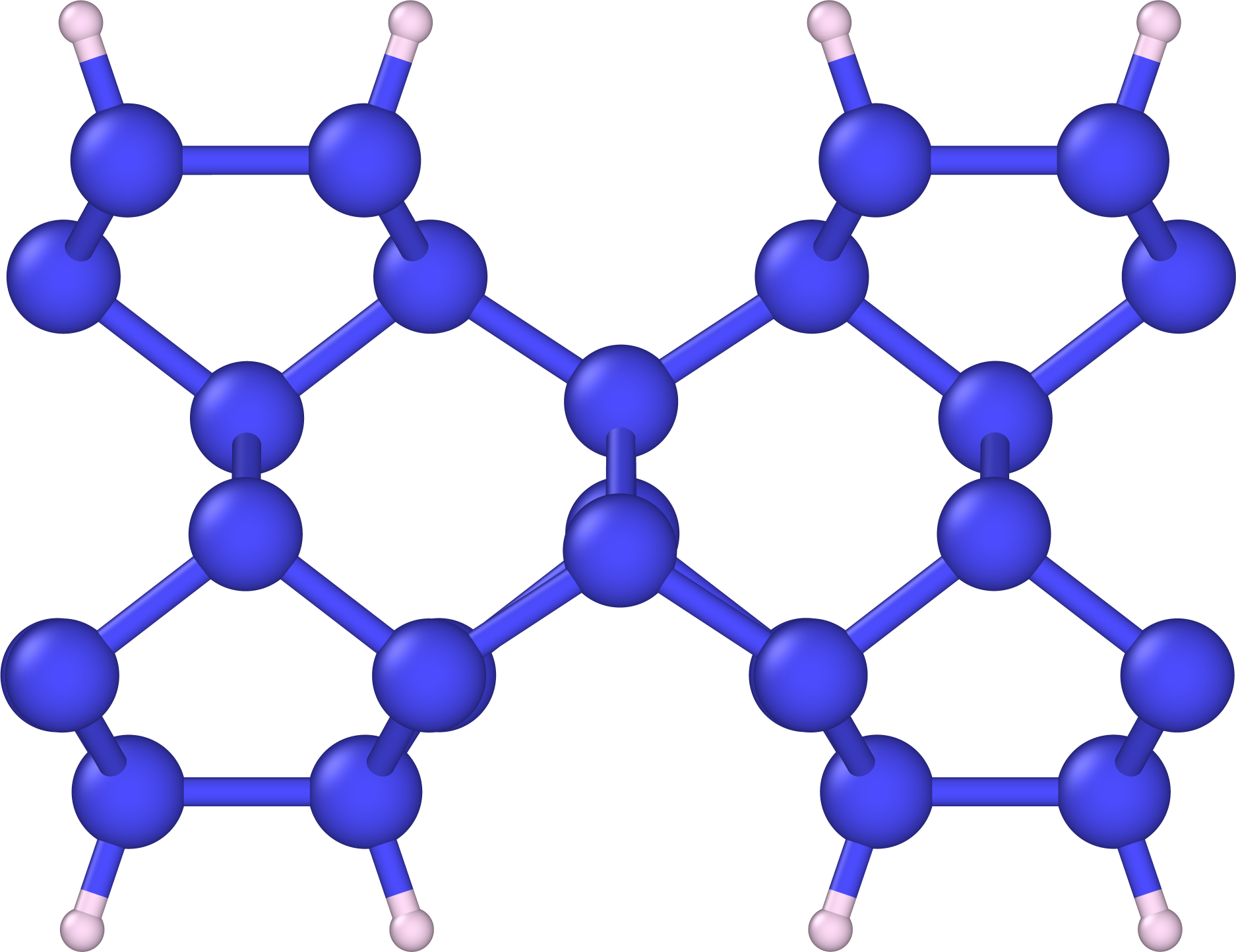}
\end{minipage}
\hspace{0.02\columnwidth}
\begin{minipage}[c][0.34\columnwidth][c]{0.45\columnwidth}
\centering
\includegraphics[width=1.15\linewidth,keepaspectratio]{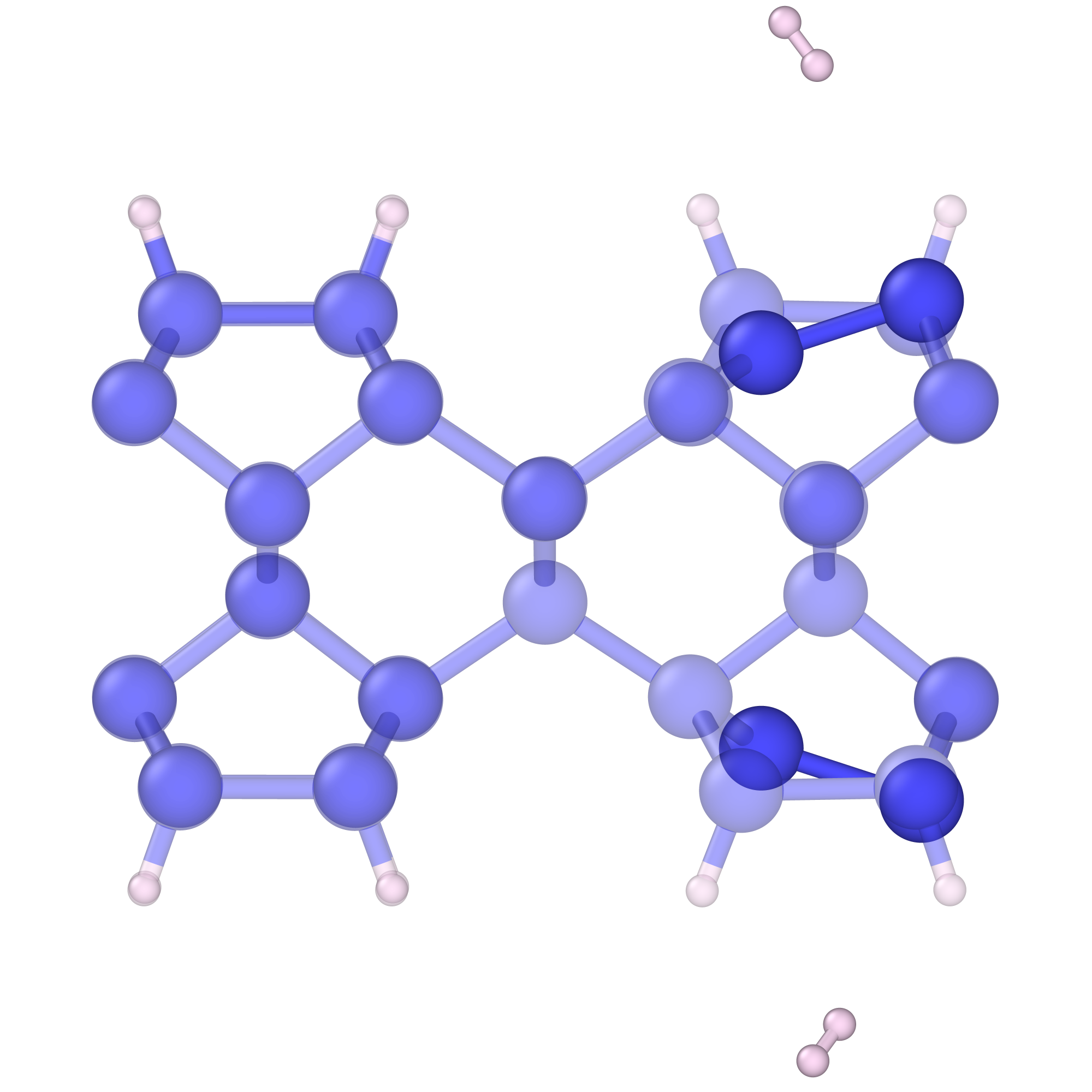}
\end{minipage}
\vspace{3em}
\caption{Si(100) surface depassivation: the initial state is the fully hydrogen-passivated surface (left), while the final state contains two surface dangling bonds and one desorbed \(\ce{H2}\) molecule per surface (right). The reacting H atoms and the Si atoms that form dangling bonds in the final state are rendered fully opaque, while the remaining atoms are shown with reduced opacity for visual clarity.}
\label{fig:si100_structures}
\end{figure}

As a flat surface reference for the same local Si--H bond-breaking chemistry, we consider a symmetric 6-layer Si(100) slab containing 48 Si atoms and hydrogen termination on both sides, as described in Sec.~\ref{sec:dft}. Before hydrogen passivation, the slab exhibits the standard dimer-reconstructed Si(100) termination. Two surface configurations are examined, see Fig.~\ref{fig:si100_structures}: (i) the fully hydrogen-passivated surface, in which each surface dangling bond is saturated by one hydrogen atom, and (ii) a depassivated configuration, in which two neighboring Si--H bonds are broken, leaving two dangling bonds and producing one \(\ce{H2}\) molecule in the vacuum region, modelled symmetrically on both surfaces. The latter corresponds to the same local reaction as Eq.~\eqref{eq:two_sih_to_h2}, but with the hydrogen molecule desorbed into vacuum (gas phase), whereas in the multivacancy case it remained confined inside the void space of the cavity.

All Si structures and hydrogen configurations considered in this work are summarized in Table~\ref{tab:configs}. These various defect configurations allow for direct comparison of the depassivation energetics between bulk defects, confined cavities, and extended surfaces, accounting for both single- and double-H processes.

\begin{table*}[t]
\caption{List of all Si structures and hydrogen configurations considered in this work, ranging from bulk interstitials, point defects, multivacancies and cavities to extended surfaces. The first column identifies the Si environment, while the second specifies the hydrogen configuration.}
\label{tab:configs}
\centering
\scriptsize
\setlength{\tabcolsep}{7pt}
\renewcommand{\arraystretch}{1.18}
\begin{tabular}{lllll}
\toprule
\textbf{Environment} & \textbf{Configuration label} & \textbf{Composition} & \textbf{DB count} & \textbf{Description} \\
\midrule
\multicolumn{5}{c}{\textbf{Bulk interstitials}} \\
Si (pristine) & H(BC) & Si$_{64}$+H & 0 & Bond-centered interstitial H \\
Si (pristine) & H(AB) & Si$_{64}$+H & 0 & Antibonding interstitial H \\
Si (pristine) & H(T$_d$)  & Si$_{64}$+H & 0 & Tetrahedral interstitial H \\
Si (pristine) & \ce{H2}(T$_d$) & Si$_{64}$+\ce{H2} & 0 & Molecular \ce{H2} at tetrahedral interstitial site \\
Si (pristine) & 2H(BC) & Si$_{64}$+2H & 0 & Two bond-centered interstitial H atoms \\
\midrule
\multicolumn{5}{c}{\textbf{Point defects}} \\
Monovacancy & V$_1$ & Si$_{63}$ & 4 & Bare monovacancy \\
Monovacancy & V$_1$H & Si$_{63}$+H & 3 & Partially passivated by one H atom \\
Monovacancy & V$_1$\ce{H2} & Si$_{63}$+2H & 2 & Partially passivated by two H atoms\\
Monovacancy & V$_1$H$_3$ & Si$_{63}$+3H & 1 & Partially passivated by three H atoms \\
Monovacancy & V$_1$H$_4$ & Si$_{63}$+4H & 0 & Fully passivated monovacancy \\
Monovacancy & V$_1$H$_3$+1H(BC) & Si$_{63}$+4H & 1 & Single-H depassivated V$_1$ with one bond-centered interstitial H \\
Monovacancy & V$_1$\ce{H2}+\ce{H2}(T$_d$) & Si$_{63}$+2H+\ce{H2} & 2 & Double-H depassivated V$_1$ with one tetrahedral interstitial \ce{H2} \\
Divacancy & V$_2$ & Si$_{62}$ & 6 & Bare divacancy \\
Divacancy & V$_2$H$_6$ & Si$_{62}$+6H & 0 & Fully passivated divacancy \\
\midrule
\multicolumn{5}{c}{\textbf{Multivacancy cavities}} \\
V$_4$ cavity & V$_4$ & Si$_{60}$ & 10 & Bare V$_4$ cavity \\
V$_4$ cavity & V$_4$H$_{10}$ & Si$_{60}$+10H & 0 & Fully passivated V$_4$ cavity \\
V$_4$ cavity & V$_4$H$_8$+\ce{H2_{conf}} & Si$_{60}$+8H+\ce{H2} & 2 & Double-H depassivated V$_4$ with trapped \ce{H2} \\
V$_6$ cavity & V$_6$ & Si$_{58}$ & 14 & Bare V$_6$ cavity \\
V$_6$ cavity & V$_6$H$_{14}$ & Si$_{58}$+14H & 0 & Fully passivated V$_6$ cavity \\
V$_6$ cavity & V$_6$H$_{12}$+\ce{H2_{conf}} & Si$_{58}$+12H+\ce{H2} & 2 & Double-H depassivated V$_6$ with trapped \ce{H2} \\
V$_6$ cavity & V$_6$H$_{13}$+H(BC) & Si$_{58}$+14H & 1 & Single-H depassivated V$_6$ with one bond-centered interstitial H \\
V$_{10}$ cavity & V$_{10}$ & Si$_{54}$ & 22 & Bare V$_{10}$ cavity \\
V$_{10}$ cavity & V$_{10}$H$_{22}$ & Si$_{54}$+22H & 0 & Fully passivated V$_{10}$ cavity \\
V$_{10}$ cavity & V$_{10}$H$_{20}$+\ce{H2_{conf}} & Si$_{54}$+20H+\ce{H2} & 2 & Double-H depassivated V$_{10}$ with trapped \ce{H2} \\
\midrule
\multicolumn{5}{c}{\textbf{Si(100) surface}} \\
Si(100) surface & Si(100)H$_{16}$ & Si$_{48}$+16H & 0 & Fully passivated monohydride symmetric Si(100) slab \\
Si(100) surface & Si(100)H$_{12}$+2\ce{H2_{vac}} & Si$_{48}$+12H+2\ce{H2} & 2/surface & \makecell[l]{\rule{0pt}{1em}Double-H depassivated two Si(100) surfaces, each with two DBs \\ and one desorbed \ce{H2}} \\
\bottomrule
\end{tabular}
\end{table*}

% ============================================================
\section{Results}
\label{sec:results}

\subsection{Thermodynamics: Stability vs \texorpdfstring{$\mu_{\mathrm H}$}{muH}}
\label{sec:thermodynamics}

% ============================================================

\begin{figure*}[t]
\centering
\includegraphics[width=1.0\textwidth]{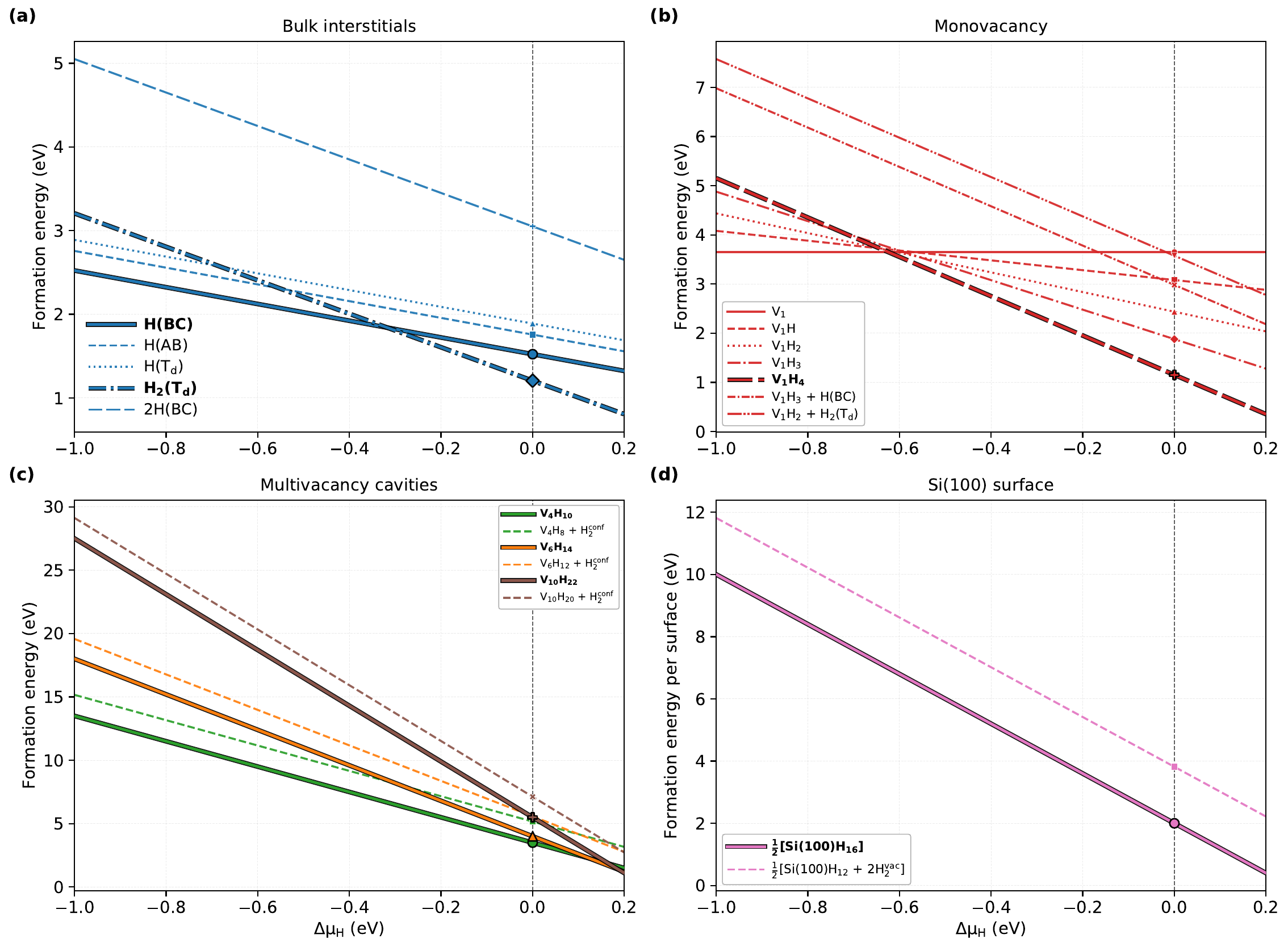}
\caption{Formation energies as a function of hydrogen chemical potential for the (a) bulk, (b) point-defects, (c) multivacancy cavities, and (d) surface configurations considered in this work. The reference \(\Delta\mu_{\mathrm H}=0\) corresponds to \(\mu_{\mathrm H}=\tfrac12E(\ce{H2})\).}
\label{fig:formation_muH}
\end{figure*}

Figure~\ref{fig:formation_muH} compares the formation energies of the bulk, monovacancy, multivacancies, and surface configurations as a function of the hydrogen chemical potential. The slope of each line is determined by the number of H atoms in the configuration. A crossing marks equal formation energies, while the lower line on either side identifies the energetically preferred configuration. Configurations with the same composition have parallel lines, whose separation is independent of \(\Delta\mu_{\mathrm{H}}\) and corresponds to their reaction energy. A direct conversion of \(\Delta\mu_{\mathrm{H}}\) to an \(\ce{H2}\) partial pressure is not made because it would require temperature-dependent vibrational and entropic contributions, which are not included in the present calculations.

\subsubsection{Bulk Hydrogen and Si Vacancy Depassivation}
\label{sec:bulk_thermo}

We begin by establishing the relative stability of different hydrogen configurations in pristine bulk silicon and at the monovacancy \(\mathrm{V_1}\) from their formation energies as a function of the hydrogen chemical potential, as shown  in Fig.~\ref{fig:formation_muH}(a) and (b). In the defect-free \(\mathrm{Si}_{64}\) supercell, the three interstitial sites for atomic hydrogen follow the energetic ordering BC \(<\) AB \(<\) T$_d$, with relative energies of \(0\,\mathrm{eV}\) (BC), \(0.234\,\mathrm{eV}\) (AB), and \(0.366\,\mathrm{eV}\) (T$_d$). More importantly, the molecular configuration in which two hydrogen atoms recombine into \(\ce{H2}\) at a tetrahedral interstitial site is stabilized relative to two isolated BC hydrogen atoms by \(\Delta E=-1.874\,\mathrm{eV}\), see Eq.~\eqref{eq:bulk_h2_reaction}. This result shows, within the present calculations, that interstitial hydrogen preferentially adopts a molecular \ce{H2} state in pristine bulk silicon, in agreement with previous first-principles studies \cite{VanDeWalle1994,Herring2001,VanDeWalle1998}.

At the Si monovacancy, sequential hydrogenation \(\mathrm{V_1H_n}\) \((n=0\text{--}4)\) exhibits a systematic decrease in formation energy with increasing hydrogen content. At the reference chemical potential \(\Delta\mu_{\mathrm H}=0\), the bare \(\mathrm{V_1}\) has a formation energy of \(3.652\,\mathrm{eV}\), while sequential hydrogenation lowers \(E_{\mathrm{f}}\) to \(1.153\,\mathrm{eV}\) for the fully passivated \(\mathrm{V_1H_4}\) complex. This identifies \(\mathrm{V_1H_4}\) as the thermodynamically preferred monovacancy configuration. As shown in Fig.~\ref{fig:formation_muH}(b), the formation energy lines of \(\mathrm{V_1}\) and \(\mathrm{V_1H_4}\) cross at \(\Delta\mu_{\mathrm{H}}\approx-0.625\,\mathrm{eV}\). Above this value, \(\mathrm{V_1H_4}\) has the lower formation energy, indicating that complete hydrogen passivation of the monovacancy is energetically favored. This trend is consistent with earlier studies of hydrogenated vacancies in crystalline silicon \cite{VanDeWalle1994DB,Kolevatov2019}.

For the single-H depassivation process of Eq.~\eqref{eq:v1_depassivation}, one Si--H bond is broken, one dangling bond is created, and the released hydrogen atom adopts a bond-centered interstitial site. The reaction energy for this final state is \(1.829\,\mathrm{eV}\). As a complementary comparison, we also considered a double-H depassivation process,
\[
\mathrm{V_1H_4}
\rightarrow
\mathrm{V_1H_2}+\ce{H2}(T_d)
\]
In this case, two Si--H bonds are simultaneously broken, and the released H atoms form one tetrahedral (T$_d$) interstitial \(\ce{H2}\) molecule.
This final state lies \(2.423\,\mathrm{eV}\) above \(\mathrm{V_1H_4}\), which is \(0.594\,\mathrm{eV}\) higher than the single-H final state \(\mathrm{V_1H_3}+\mathrm{H(BC)}\). Thus, for monovacancy depassivation, the single-H process with an atomic H(BC) final state is energetically preferred over the double-H process with a molecular \(\ce{H2}(T_d)\).

\subsubsection{Multivacancy Depassivation and Cavity-Confined \ce{H2}}
\label{sec:void_thermo}

To compare (de)passivation of point defects to the case of larger defect structures, we considered connected multivacancy clusters \(\mathrm{V_4}\), \(\mathrm{V_6}\), and \(\mathrm{V_{10}}\). For each cluster, the initial state is the fully passivated cavity-wall configuration in which all available dangling bonds are terminated by hydrogen atoms, giving \(\mathrm{V_4H_{10}}\), \(\mathrm{V_6H_{14}}\), and \(\mathrm{V_{10}H_{22}}\). The final state is the corresponding double-H depassivated configuration with one confined \(\ce{H2}\) molecule, as defined in Sec.~\ref{sec:multivacancy_configurations}. These initial-state and final-state configurations are compared in Fig.~\ref{fig:formation_muH}(c). The reaction energies are \(1.663\,\mathrm{eV}\), \(1.574\,\mathrm{eV}\), and \(1.621\,\mathrm{eV}\) for \(\mathrm{V_4}\), \(\mathrm{V_6}\), and \(\mathrm{V_{10}}\), respectively. In all three cases, the depassivated state remains higher in energy than the fully passivated starting point. The relaxed final structures contain a confined \(\ce{H2}\) molecule inside the cavity, rather than a bulk interstitial \ce{H2} state as considered for the monovacancy \(\mathrm{V_1}\) case. This explains the significantly lower energy cost of the double-H process for the multivacancies in comparison to the energy cost of \(2.423\,\mathrm{eV}\) for the monovacancy double-H process. Within the set of cavities examined here, \(\mathrm{V_6}\) gives the lowest double-H depassivation energy. Previous first-principles work identified \(\mathrm{V_6}\) as an exceptionally stable Si multivacancy and showed that hydrogen-decorated multivacancies can trap molecular \(\ce{H2}\) \cite{AkiyamaOshiyama2001}. The final state considered here is different because the confined \(\ce{H2}\) forms through the depassivation of two Si--H bonds in the presence of two resulting dangling bonds on the cavity wall. The present study therefore extends the previous work to direct depassivation and the formation of confined \(\ce{H2}\) in a partially passivated cavity. Experimental and theoretical studies have also associated molecular \(\ce{H2}\) with void-like or multivacancy-related defects in Si \cite{Hourahine1999,Ishioka1999,Mori2001,Ghasemi2014}. 

For comparison, we also constructed single-H depassivated \(\mathrm{V_6H_{13}}+\mathrm{H(BC)} \) configuration, analogous to the \(\mathrm{V_1H_3}+\mathrm{H(BC)}\) final state for the monovacancy. Its reaction energy is \(1.955\,\mathrm{eV}\), which is \(0.381\,\mathrm{eV}\) higher than the \(1.574\,\mathrm{eV}\) obtained for double-H depassivated state. The relaxed structures also show that the dangling bond is located on different Si atoms under different doping conditions. Further structural details are provided in Appendix~\ref{app:v6_single_h}. The following kinetic analysis therefore focuses on the double-H pathway.

\subsubsection{Si(100) Surface Depassivation}
\label{sec:surface_thermo}

The same Si--H bond-breaking chemistry can be examined on the Si(100) surface. For the symmetric Si(100) slab, we compared a fully hydrogenated surface with a depassivated configuration in which, for each of the two surfaces, two neighboring Si--H bonds are broken, leaving two dangling bonds and producing one \(\ce{H2}\) molecule in the vacuum region. This process again corresponds to reaction~\eqref{eq:two_sih_to_h2}, but with the hydrogen molecule desorbed into vacuum (or gas phase) rather than remaining confined at a bulk interstitial site or inside a bulk cavity. Because the two configurations have the same composition, they appear as parallel lines in Fig.~\ref{fig:formation_muH}(d), with the depassivated state lying \(1.816\,\mathrm{eV}\) per surface above the fully passivated state.

Compared to the \(\mathrm{V_4}\), \(\mathrm{V_6}\), and \(\mathrm{V_{10}}\) multivacancy results, the Si(100) calculation provides an extended-surface limit for the same double Si--H bond-breaking chemistry. The largest cavity considered here was \(\mathrm{V_{10}}\); in principle, cavities of increasing size can be viewed as eventually approaching this surface limit. For the multivacancy cavities, depassivation energies in the range \(1.574\text{--}1.663\,\mathrm{eV}\) were found, slightly below the corresponding Si(100) surface value of \(1.816\,\mathrm{eV}\). This comparison shows that the reaction is energetically similar at confined cavity walls and in the surface limit. The slightly more favourable energy for the confined processes might be due to a residual stabilizing interaction between the cavity-confined \ce{H2} and the dangling bonds of the cavity wall.

%\begin{widetext}
%\begin{center}
\begin{table*}[t]
\caption{Reaction energies for bulk \(\ce{H2}\) formation and Si--H depassivation reactions. In each reaction, the initial and final states contain the same number of Si and H atoms, so \(\Delta E\) is calculated directly from Eq.~\eqref{eq:reaction_energy}.}
\label{tab:thermo_summary}
\small
\setlength{\tabcolsep}{5.0pt}
\renewcommand{\arraystretch}{1.18}
\vspace{-0.1cm}
\begin{tabular}{llll}
\toprule
\textbf{Environment} & \textbf{Initial state} & \textbf{Final state} & \(\boldsymbol{\Delta E}\) \\
\midrule
Bulk Si 
    & \(2\mathrm{H(BC)}\)
    & \(\ce{H2}(T_d)\)
    & \(-1.874\,\mathrm{eV}\) \\
Monovacancy
    & \(\mathrm{V_1H_4}\)
    & \(\mathrm{V_1H_3} + \mathrm{H(BC)}\)
    & \(1.829\,\mathrm{eV}\) \\
Monovacancy
    & \(\mathrm{V_1H_4}\)
    & \(\mathrm{V_1H_2} + \ce{H2}(T_d)\)
    & \(2.423\,\mathrm{eV}\) \\
Cavity wall \(\mathrm{V_4}\)
    & \(\mathrm{V_4H_{10}}\)
    & \(\mathrm{V_4H_8} + \ce{H2_{conf}}\)
    & \(1.663\,\mathrm{eV}\) \\
Cavity wall \(\mathrm{V_6}\)
    & \(\mathrm{V_6H_{14}}\)
    & \(\mathrm{V_6H_{12}} + \ce{H2_{conf}}\)
    & \(1.574\,\mathrm{eV}\) \\
Cavity wall \(\mathrm{V_{10}}\)
    & \(\mathrm{V_{10}H_{22}}\)
    & \(\mathrm{V_{10}H_{20}} + \ce{H2_{conf}}\)
    & \(1.621\,\mathrm{eV}\) \\
Si(100) surface 
    & \(\mathrm{Si(100)H_{16}}\)
    & \(\mathrm{Si(100)H_{12}} + 2\ce{H2_{vac}}\)
    & \(1.816\,\mathrm{eV}\) per surface \\
\bottomrule
\end{tabular}
%\vspace{-0.1cm}
%\end{center}
%\end{widetext}
\end{table*}

Table~\ref{tab:thermo_summary} summarizes the main reaction energies. Most notably, the reaction energy of a single-H depassivation process was found to be higher than the reaction energy of double-H processes with the formation of cavity-confined \ce{H2}. Although the cavity reactions break two Si--H bonds, their reaction energies are \(1.574\text{--}1.663\,\mathrm{eV}\), below the \(1.829\,\mathrm{eV}\) needed for monovacancy single-H depassivation. This indicates that confined \(\ce{H2}\) formation overcompensates for the additional cost of breaking a second Si--H bond.

\subsubsection{Electronic Structure of the Single-H and Double-H Final States}

\begin{figure*}[t]
    \centering
    \includegraphics[width=\textwidth]{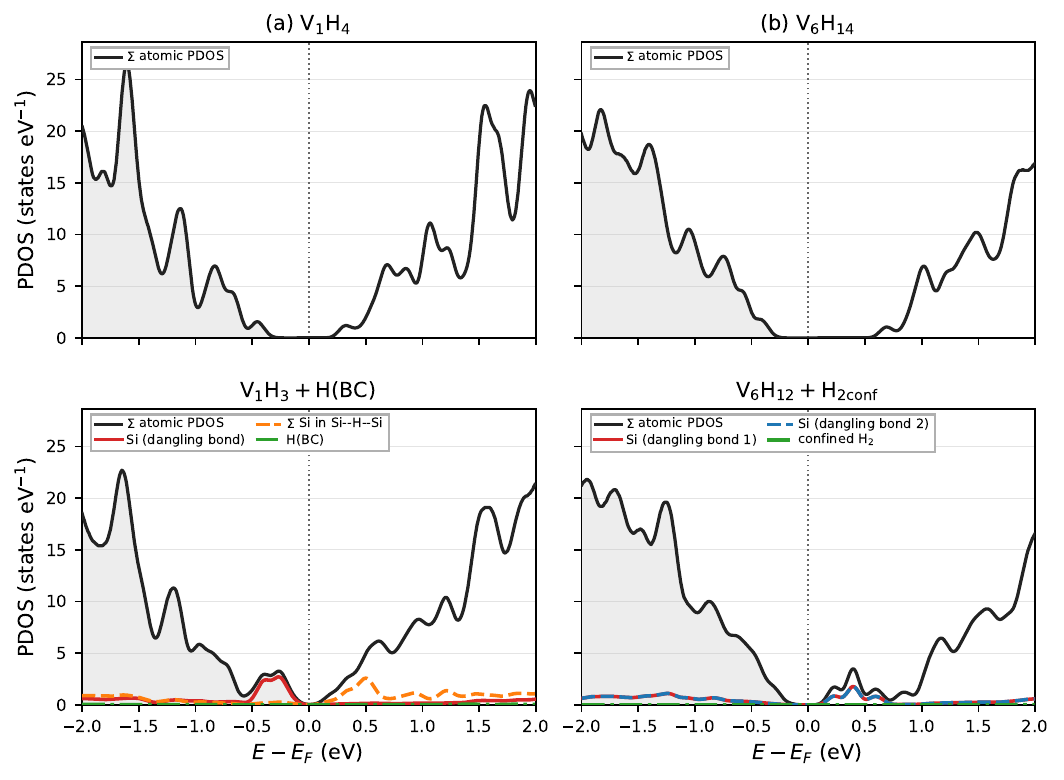}
    \caption{PBE projected densities of states for the (a) single-H and (b) double-H depassivation processes. The upper panels show the fully passivated initial states, and the lower panels show the corresponding depassivated final states. The vertical dashed line denotes the Fermi level \(\mathrm{E_F}\).}
    \label{fig:pdos_pbe}
\end{figure*}

To determine how Si--H bond breaking modifies the electronic structure, Fig.~\ref{fig:pdos_pbe} shows the PBE projected densities of states before and after the single-H and double-H depassivation processes. The fully passivated initial states are shown as references and exhibit clean band gaps, while the corresponding depassivated \(\mathrm{V_1H_3+H(BC)}\) and \(\mathrm{V_6H_{12}+H_{2\mathrm{conf}}}\) final states introduce electronic states. Here, HOMO and LUMO denote the highest occupied and lowest unoccupied Kohn--Sham states of the calculated supercell, respectively.

For \(\mathrm{V_1H_3+H(BC)}\), the HOMO has its largest contribution from the Si atom where the dangling bond is created, whereas the LUMO has its largest contribution from the two Si atoms forming the Si--H--Si bond-centered configuration. The calculated occupations are 0.9995 and 0.0006 per spin for the HOMO and LUMO, respectively. The HOMO is therefore fully occupied rather than half-filled, and the system retains a residual band gap of \(0.204\,\mathrm{eV}\).

For \(\mathrm{V_6H_{12}+H_{2\mathrm{conf}}}\), neither of the two Si atoms where the dangling bonds are created contributes prominently to the HOMO. These two Si atoms make the largest and nearly equal contributions to the LUMO. The corresponding residual band gap is \(0.380\,\mathrm{eV}\).

The Löwdin charge analysis provides additional insight into how electron density is redistributed among the atoms involved in the single-H and double-H final states. Using the definition employed by Diggs et al. \cite{Diggs2025FermiLevel}, the effective charge of Si atom $i$ is defined as 
\begin{equation}
    q_i = \overline{N}_{\mathrm{Si},j\ne i}-N_i,
\end{equation}
where $N_i$ is the Löwdin population of atom $i$ and $\overline{N}_{\mathrm{Si},j\ne i}$ is the average Löwdin population of the other Si atoms. For H, $q_H=1-N_H$. The effective charges associated with H(BC) and the confined $\mathrm{H_2}$ molecule are calculated as
\begin{align}
    q_{H(BC)} &= q_{\mathrm{BC1}}+q_{\mathrm{BC2}}+q_H, \\
    q_{\mathrm{H_2}} &= q_{\mathrm{H(1)}}+q_{\mathrm{H(2)}}.
\end{align}
Negative values indicate increased electronic population, whereas positive values indicate reduced electron density. 

\begin{table}[H]
\centering
\caption{PBE Löwdin effective charges for the
$\mathrm{V_1H_3+H(BC)}$ single-H final state.}
\label{tab:lowdin_v1}
\begin{tabular}{lc}
\hline\hline
Quantity & Effective charge ($e$) \\
\hline
$q_{\mathrm{DB}}$    & $-0.206$ \\
$q_{\mathrm{BC1}}$   & $+0.197$ \\
$q_{\mathrm{BC2}}$   & $+0.188$ \\
$q_{H(BC)}$          & $+0.281$ \\
$q_H$                & $-0.104$ \\
\hline\hline
\end{tabular}
\end{table}

The negative $q_{\mathrm{DB}}$ indicates increased electron density on the Si atom where the dangling bond is created, whereas the positive $q_{H(BC)}$ shows an overall reduction in electron density associated with the H(BC) complex. This reduction comes mainly from the two Si atoms forming the Si--H--Si bond-centered configuration. Together with the HOMO and LUMO contributions and occupations, this charge pattern is consistent with a redistribution of electronic populations from the higher-energy H(BC) state to the lower-energy dangling-bond state.

This is consistent with Diggs et al. \cite{Diggs2025FermiLevel}, who studied the same single-H depassivation process and found that the Si dangling bond state was occupied by two electrons. They attributed this occupation to electron donation from the higher-energy H(BC) state to the lower-energy dangling-bond state. They reported $q_{\mathrm{DB}}=-0.177\,e$, $q_{\mathrm{BC1}}=+0.184\,e$, $q_{\mathrm{BC2}}=+0.192\,e$, and $q_{H(BC)}=+0.338\,e$. The similar charge pattern, with increased electron population on the dangling-bond Si and depletion on the two Si atoms forming the Si--H--Si configuration, supports the same interpretation of electronic redistribution for the present single-H final state.

For \(\mathrm{V_6H_{12}+H_{2\mathrm{conf}}}\), the two Si atoms where the dangling bonds are created each have $q_{\mathrm{DB}}=-0.038\,e$, whereas the confined \(\ce{H2}\) molecule has a combined effective charge of $q_{\mathrm{H_2}}=+0.037\,e$. These values remain much closer to zero than those obtained for the single-H final state. Therefore, the Löwdin populations do not show electronic redistribution of comparable magnitude in the double-H final state. Consistently, the PDOS in Fig.~\ref{fig:pdos_pbe} places the Si dangling-bond character in the empty LUMO rather than the occupied HOMO.

The corresponding HSE06 electronic-structure analysis is discussed in Appendix~\ref{app:pdos}.
The kinetic barriers for these processes are examined next.
% ============================================================
\subsection{Kinetics of De- and Repassivation}
\label{sec:kinetics}
% ============================================================

\begin{figure*}[t]
\centering
\includegraphics[width=1.0\textwidth]{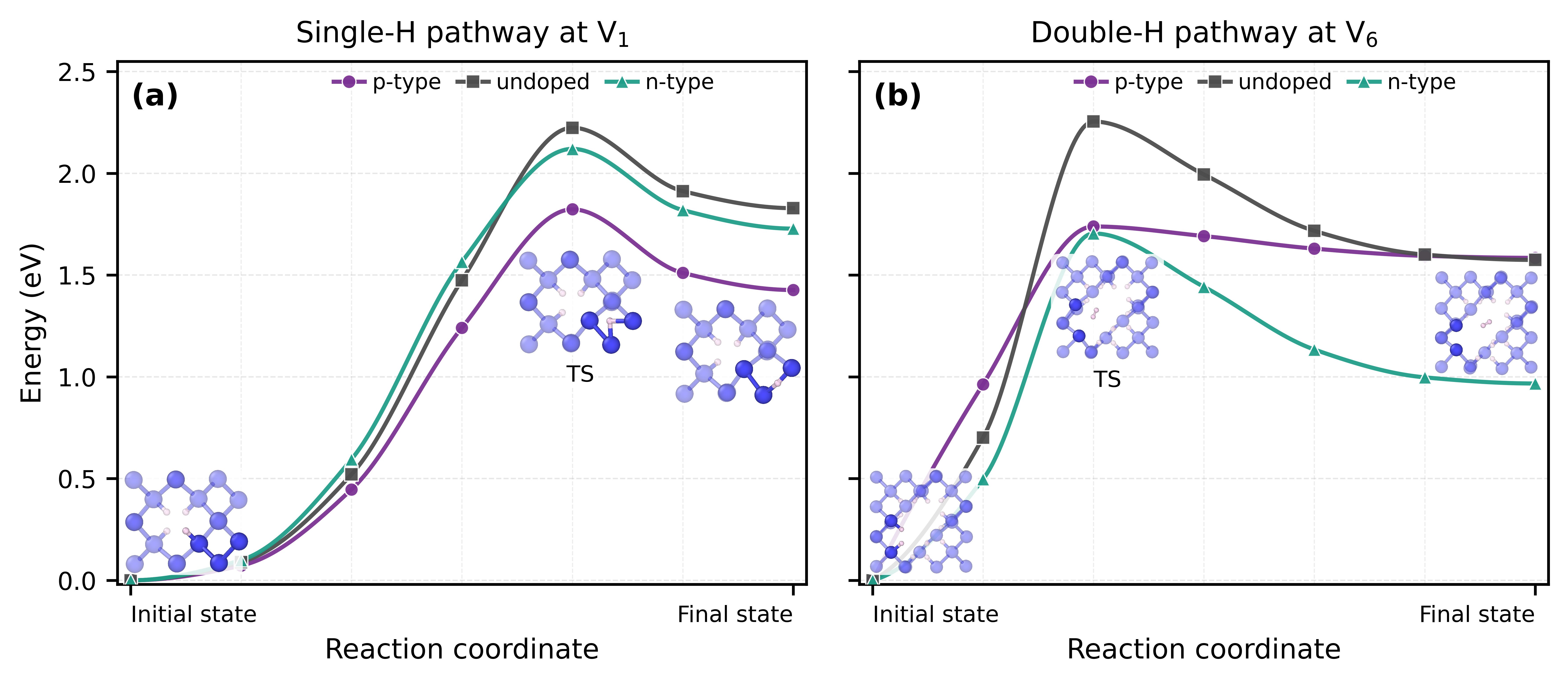}
\caption{Energy profiles for (a) the single-H pathway in \(\mathrm{V_1}\), corresponding to Eq.~\eqref{eq:v1_depassivation} and (b) the double-H pathway \(\mathrm{V_6}\), corresponding to Eq.~\eqref{eq:two_sih_to_h2}, obtained from NEB calculations at the PBE level of theory. The profiles are shown under undoped, n-type, and p-type conditions.}
\label{fig:neb_v1_v6}
\end{figure*}

The thermodynamic analysis identified the relevant passivated and depassivated states for each structural environment. As shown in Table~\ref{tab:thermo_summary}, the reaction energies motivate a direct comparison of the corresponding activation barriers. To test whether the lower reaction energy of the double-H process relative to the single-H process is also corroborated by the kinetics, PBE minimum-energy paths were computed using the NEB method for the monovacancy single-H reaction and the \(\mathrm{V_6}\) double-H reaction. The resulting forward and reverse barriers are summarized in Table~\ref{tab:pbe_barriers}, and the corresponding NEB energy profiles are shown in Fig.~\ref{fig:neb_v1_v6}.

\subsubsection{Single-H depassivation at the monovacancy}
\label{sec:v1_neb}

Figure~\ref{fig:neb_v1_v6} (a) shows the NEB profiles for monovacancy single-H depassivation. The corresponding barriers are listed in Table~\ref{tab:pbe_barriers}, with the lowest forward barrier obtained in the p-type case. The numerical values agree closely with the monovacancy Si--H bond breaking barriers reported by Diggs et al. for the same depassivated final state, in which the released H atom occupies a bond-centered interstitial site \cite{DiggsThesis2025,Diggs2025FermiLevel}. In their study, the lowest forward barrier was also obtained in the p-type case, with \(E_{\mathrm a}^{\rightarrow}=1.73\,\mathrm{eV}\), compared with \(2.07\,\mathrm{eV}\) for intrinsic and \(1.98\,\mathrm{eV}\) for n-type.

\subsubsection{Double-H cavity depassivation with direct \texorpdfstring{\(\ce{H2}\)}{H2} formation}
\label{sec:v6_neb}

Figure~\ref{fig:neb_v1_v6} (b) shows the NEB profiles for double-H depassivation of the \(\mathrm{V_6}\) cavity with direct formation of confined \(\ce{H2}\). Compared with the monovacancy pathway, the \(\mathrm{V_6}\) forward barriers are similar in the undoped case but significantly lower for both n-type and p-type results. The reverse barrier is particularly small in the p-type case \(\approx 0.154\,\mathrm{eV}\), indicating that local repassivation via splitting of confined \(\ce{H2}\) can be achieved without major activation once the molecule is present near two cavity-wall dangling bonds.
This comparison highlights a key kinetic finding. Although the double-H pathway breaks two Si--H bonds, its forward barrier is comparable to, or even lower than that of the single-H pathway, which involves the rupture of only one Si--H bond. This indicates that the formation of the H--H bond in confined \(\ce{H2}\) overcompensates for the energetic cost of breaking the second Si--H bond. Together with the reaction energies in Table~\ref{tab:thermo_summary}, these barriers show that confined \(\ce{H2}\) molecules are not only stored in the cavity, but also participate directly in local depassivation and repassivation. 

\begin{table}[H]
\centering
\caption{PBE activation barriers for (a) single-H point-defect
depassivation and (b) double-H cavity depassivation. The arrows indicate
depassivation \((\rightarrow)\) and repassivation \((\leftarrow)\).}
\label{tab:pbe_barriers}

\small
\setlength{\tabcolsep}{7pt}
\renewcommand{\arraystretch}{1.18}

\begin{tabular}{lcc}
\toprule
\textbf{Doping}
& \(\boldsymbol{E_{\mathrm a}^{\rightarrow}}\) (eV)
& \(\boldsymbol{E_{\mathrm a}^{\leftarrow}}\) (eV) \\
\midrule

\multicolumn{3}{l}{
\textbf{(a) Single-H pathway}
} \\
\multicolumn{3}{l}{
\(\mathrm{V_1H_4}\longleftrightarrow
\mathrm{V_1H_3}+\mathrm{H(BC)}\)
} \\
\addlinespace[2pt]

Undoped & 2.224 & 0.395 \\
\(n\)-type & 2.120 & 0.392 \\
\(p\)-type & 1.823 & 0.396 \\

\midrule

\multicolumn{3}{l}{
\textbf{(b) Double-H pathway}
} \\
\multicolumn{3}{l}{
\(\mathrm{V_6H_{14}}\longleftrightarrow
\mathrm{V_6H_{12}}+\ce{H2_{conf}}\)
} \\
\addlinespace[2pt]

Undoped & 2.254 & 0.680 \\
\(n\)-type & 1.705 & 0.737 \\
\(p\)-type & 1.739 & 0.154 \\

\bottomrule
\end{tabular}
\end{table}

%\FloatBarrier

\subsection{Hybrid functional validation}
\label{sec:hse_validation}

To further assess the robustness of the PBE results, the key kinetic comparison was checked with HSE06 single-point calculations on the corresponding PBE initial state, transition state, and final state geometries. These calculations are not fully optimized HSE06 NEB pathways, but they were used to test whether the main single-H versus double-H comparisons are robust against common GGA-level errors, including the underestimated band gap, self-interaction errors, and the resulting delocalization of electronic states.

The PBE calculations were performed using projector-augmented-wave (PAW) pseudopotentials, whereas norm-conserving (NC) pseudopotentials were used for the HSE06 single-point calculations due to the large memory requirements of hybrid-functional calculations with PAW pseudopotentials. Additional PBE single-point calculations with the same NC pseudopotentials were therefore performed to separate the effect of the pseudopotential type from that of the exchange-correlation functional.

For HSE06, the fraction of exact exchange was set to the default value of \(0.25\), corresponding to \(25\%\) Hartree--Fock exchange, and the screening parameter was set to \(0.106\,\mathrm{bohr^{-1}}\), which determines the range of the short-range exact-exchange interaction.

\begin{figure}[t]
\centering
\includegraphics[width=0.825\columnwidth]{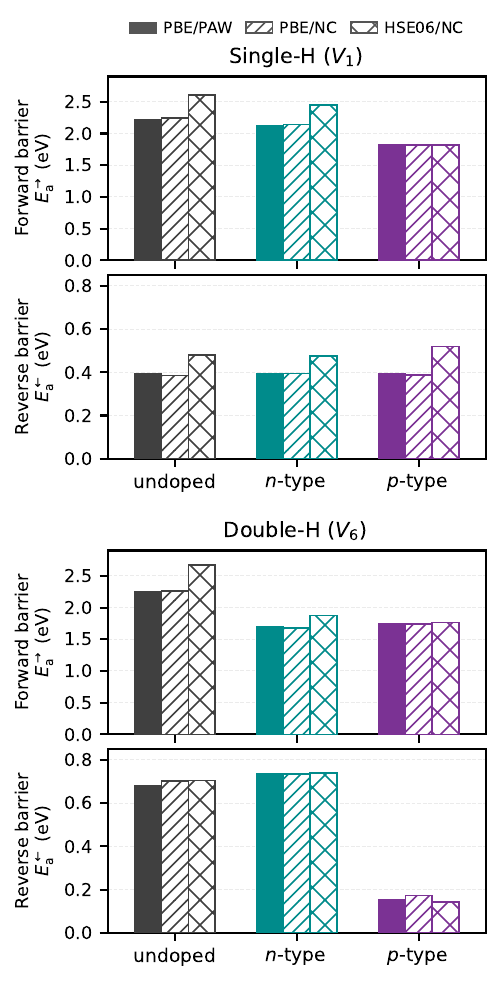}
\caption{Bar-chart comparison of the forward and reverse activation barrier estimates for the single-H and double-H processes obtained under different doping conditions.}
\label{fig:pbe_vs_hse_fig}
\end{figure}

The HSE06 results support the PBE picture in two specific ways, as summarized in Table~\ref{tab:hse_barriers} and shown in Fig.~\ref{fig:pbe_vs_hse_fig}. First, the direct double-H cavity pathway remains kinetically competitive with the single-H monovacancy pathway. It is lower in both n-type and p-type cases. Second, the low reverse barrier for the p-type double-H pathway is preserved at the hybrid-functional level, with an HSE06 estimate of \(E_{\mathrm a}^{\leftarrow}=0.142\,\mathrm{eV}\).

\begin{table*}[t]
%\begin{widetext}
%\begin{center}
    \centering
    \caption{Forward and reverse activation barrier estimates for the single-H and double-H processes obtained with PBE/PAW, PBE/NC, and HSE06/NC.}
    \label{tab:hse_barriers}
    \small
    \setlength{\tabcolsep}{6pt}
    \renewcommand{\arraystretch}{1.18}
    \begin{tabular}{llccc}
    \toprule
    \textbf{Process}
    & \textbf{Carrier condition}
    & \textbf{Method}
    & \(\boldsymbol{E_{\mathrm a}^{\rightarrow}}\) (eV)
    & \(\boldsymbol{E_{\mathrm a}^{\leftarrow}}\) (eV)
    \\
    \midrule
    \multirow{9}{*}{Single-H \((\mathrm{V_1}\))} 
    & \multirow{3}{*}{undoped}
    & PBE/PAW  & 2.224 & 0.395 \\
    &
    & PBE/NC   & 2.243 & 0.385 \\
    &
    & HSE06/NC & 2.610 & 0.480 \\
    \cmidrule(lr){2-5}
    & \multirow{3}{*}{\(n\)-type}
    & PBE/PAW  & 2.120 & 0.392 \\
    &
    & PBE/NC   & 2.141 & 0.395 \\
    &
    & HSE06/NC & 2.453 & 0.475 \\
    \cmidrule(lr){2-5}

    & \multirow{3}{*}{\(p\)-type}
    & PBE/PAW  & 1.823 & 0.396 \\
    &
    & PBE/NC   & 1.819 & 0.389 \\
    &
    & HSE06/NC & 1.824 & 0.520 \\
    \midrule
    
    \multirow{9}{*}{Double-H (\(\mathrm{V_6}\))}
    & \multirow{3}{*}{undoped}
    & PBE/PAW  & 2.254 & 0.680 \\
    &
    & PBE/NC   & 2.259 & 0.703 \\
    &
    & HSE06/NC & 2.671 & 0.704 \\
    \cmidrule(lr){2-5}
    
    & \multirow{3}{*}{\(n\)-type}
    & PBE/PAW  & 1.705 & 0.737 \\
    &
    & PBE/NC   & 1.681 & 0.736 \\
    &
    & HSE06/NC & 1.877 & 0.738 \\
    \cmidrule(lr){2-5}
    
    & \multirow{3}{*}{\(p\)-type}
    & PBE/PAW  & 1.739 & 0.154 \\
    &
    & PBE/NC   & 1.743 & 0.173 \\
    &
    & HSE06/NC & 1.767 & 0.142 \\
    \bottomrule
    \end{tabular}
\end{table*}
%\end{center}
%\end{widetext}

\section{Discussion}
\label{sec:discussion}
Previous first-principles studies have established molecular \(\ce{H2}\) in silicon as a stable interstitial species in bulk Si or as a trapped species in vacancy aggregates and void-like structures \cite{VanDeWalle1994,VanDeWalle1998,VanDeWalleNeugebauer2006,AkiyamaOshiyama2001,Hourahine1999,Ishioka1999,Mori2001}. In parallel, transition-state studies of Si--H bond breaking have mainly focused on single-H depassivation pathways, where breaking one Si--H bond produces a dangling bond and an atomic-H configuration \cite{DiggsThesis2025,Diggs2023HydrogenDegradation,Diggs2025FermiLevel}. The present work connects these two lines of research by treating molecular \(\ce{H2}\) not only as a stored species, but as part of the local Si--H bond breaking and bond reforming pathway.

\subsection{Implications for Silicon Heterojunction and Photovoltaic Interfaces}
The present results directly address hydrogen-rich regions in silicon where dangling bonds, free volume, and molecular hydrogen can coexist. This situation is expected to occur near amorphous/crystalline interfaces in silicon heterojunction cells, where porous regions in underdense hydrogenated amorphous silicon can host molecular \(\ce{H2}\) \cite{Liu2016Underdense,Fischer2023}. In these environments, passivation is unlikely to be a purely static property of isolated Si--H bonds. Instead light soaking and illuminated annealing suggest that passivation can involve local bond breaking, hydrogen rearrangement, and bond reformation \cite{Mahtani2013,Kobayashi2016APL,Hammann2023JPHOTOV}.

The atomistic mechanism identified here provides one possible step in this process. The reverse barrier exhibits a pronounced dependence on the doping condition, decreasing from \(0.680\,\mathrm{eV}\) in the undoped case and \(0.737\,\mathrm{eV}\) in the n-type case to \(0.154\,\mathrm{eV}\) in the p-type case. The particularly low p-type barrier indicates that, once confined \(\mathrm{H_2}\) is already positioned near the two dangling bonds, its dissociation and the reformation of the two Si--H bonds require little additional activation. This local reaction is therefore predicted to be thermally accessible at room temperature even without illumination. It consequently cannot explain why illumination or heating is required for the overall recovery process. However, experimental studies support a distinction between this local reaction and the preceding hydrogen processes. Studies of p-type crystalline silicon during elevated-temperature dark annealing found changes in hydrogen configurations and defect activity without illumination \cite{Hammann2023JPHOTOV,Hammann2024SolarRRL}. These studies do not identify the local Si--H reformation considered here, but show that illumination is not required for all hydrogen-related reactions. The need for elevated temperature in those experiments suggests that activation may be required during preceding hydrogen release or transport steps in the recovery process. Kwapil et al. observed that illumination reduced the concentration of boron-hydrogen pairs at \(25\,^{\circ}\mathrm{C}\). They explained this decrease by illumination-generated electrons helping hydrogen detach from boron, thereby increasing the amount of atomic hydrogen available for subsequent reactions \cite{Kwapil2026LightSoaking}. Fischer et al. found that the time required for passivation to improve during intense light soaking is mainly determined by hydrogen redistribution and changes in hydrogen bonding configurations in hydrogenated amorphous silicon \cite{Fischer2025CellReports}. They also related hydrogen transport in underdense hydrogenated amorphous silicon to molecular \(\mathrm{H_2}\) diffusion through a void network \cite{Fischer2023}. Illumination and annealing may therefore control hydrogen redistribution and transport to the defect, whereas the calculated reverse barrier describes the subsequent local repassivation step, after molecular \(\mathrm{H_2}\) has reached the defect. 

The relation to the doping character of silicon heterojunction layers is qualitative. Kobayashi et al. found that light soaking improved passivation for structures containing either p-type or n-type doped hydrogenated amorphous silicon, whereas a structure containing only intrinsic hydrogenated amorphous silicon degraded \cite{Kobayashi2016APL}. The experimental effect is therefore not restricted to p-type layers. The lower barrier in the p-type case indicates that the final local reaction is more accessible, whereas the undoped and n-type calculations indicate a comparatively larger activation requirement for the same local reaction.

The present crystalline vacancy model does not describe the full SHJ interface and includes neither amorphous disorder, explicit illumination, nor nonequilibrium carriers. Therefore, the results should be interpreted as a local reaction pathway that may contribute to understanding illuminated annealing recovery and not as a device-level model.

\subsection{Local barrier scales and rate estimates}
The activation barriers can be used to estimate local reaction times within a simplified version of transition state theory, using Arrhenius expression, \(k\approx\nu\exp[-E_{\mathrm a}/(k_{\mathrm B}T)]\). The corresponding local barrier-crossing time is estimated from the inverse rate constant, \(\tau=k^{-1}\), giving \(\tau \approx \nu^{-1}\exp(E_{\mathrm a}/k_{\mathrm B}T)\), using an attempt frequency of \(\nu \approx 10^{13}\,\mathrm{s^{-1}}\) \cite{Hwang2006SiDiff}. These estimated times should only be taken as a rough indication of whether one calculated local reaction is faster or slower than another. They are not predictions of device degradation or recovery rates, because the total energies calculated in this work are purely DFT energies and do not include zero-point, vibrational or finite-temperature corrections. Long-range effects such as hydrogen diffusion and defect concentrations are also not included. 

For experimental context, Fischer et al.\ reported that passivation improvement under intense light soaking at \(250\,^{\circ}\mathrm{C}\) was observed on time scales as short as \(\approx 0.2\,\mathrm{s}\) \cite{Fischer2025CellReports}. At this temperature, the calculated reverse barriers of  \(0.680\), \(0.737\), and \(0.154\,\mathrm{eV}\) for the undoped, n-type, and p-type double-H pathways correspond, within the same Arrhenius estimate, to barrier-crossing times of approximately \(3.6\times10^{-7}\), \(1.3\times10^{-6}\), and
\(3.0\times10^{-12}\,\mathrm{s}\), respectively. Thus, even the longest estimated barrier-crossing time, obtained for the n-type case, is approximately five orders of magnitude faster than the experimentally observed light-soaking time scale. 

For comparison with hydrogen processes at higher temperatures, the estimated time scales were evaluated at \(700\,\mathrm{K}\), which is close to a hydrogen-effusion feature observed in underdense a-Si:H near \(400\,^{\circ}\mathrm{C}\), which has been associated with molecular \(\ce{H2}\) diffusion through a void network \cite{Fischer2023}. At this temperature, the undoped PBE \(\mathrm{V_6}\) depassivation barrier of \(2.254\,\mathrm{eV}\) gives an approximate time of \(1.7\times10^{3}\,\mathrm{s}\), whereas the reverse barrier of \(0.680\,\mathrm{eV}\) gives \(7.9\times10^{-9}\,\mathrm{s}\). For the p-type \(\mathrm{V_6}\) pathway, the reverse barrier is smaller \(\approx 0.154\,\mathrm{eV}\) at the PBE level and \(0.142\,\mathrm{eV}\) in the HSE06 single-point estimate, corresponding to times on the order of \(10^{-12}\,\mathrm{s}\). 

These comparisons support the interpretation that once confined \(\ce{H2}\) has reached the vicinity of dangling bonds, the final repassivation step can proceed very rapidly, with the shortest estimated barrier-crossing time obtained for p-type c-Si. Importantly, the calculated repassivation times characterize only this final passivation step and should not be interpreted as device-level recovery times.

\section{Conclusion}
\label{sec:conclusion}

This work shows that confined molecular \(\ce{H2}\) has an important role in direct Si--H depassivation and repassivation in defective silicon. In particular, double-H processes can be competitive with single-H processes. Although the single-H and double-H pathways were studied here for the specific model environments of \(\mathrm{V_1}\) and \(\mathrm{V_6}\), respectively, the main distinction is the number of Si--H bonds involved, and similar energetic comparisons can be expected in other defect environments.

The main result is that confined \(\ce{H2}\) is not only relevant as a molecular hydrogen reservoir. In defect environments with sufficient free volume, it can be the direct product of Si--H depassivation and the direct reactant for repassivation. Despite involving the breaking of two Si--H bonds, the double-H pathway in \(\mathrm{V_6}\) has a forward barrier comparable to the single-H pathway in \(\mathrm{V_1}\) in the undoped case, and even lower in the n-type and p-type cases. This indicates that the formation of the H--H bond can overcompensate for the additional cost of breaking a second Si--H bond. The calculated barriers also show a clear doping dependence. In the p-type case, the reverse barrier for the \(\mathrm{V_6}\) double-H pathway is \(0.154\,\mathrm{eV}\) at the PBE level and \(0.142\,\mathrm{eV}\) from the HSE06 estimate, suggesting that confined \(\ce{H2}\) can enable fast local recovery of Si--H passivation.

The low recovery barrier supports a dynamic picture of hydrogen passivation in underdense silicon regions, relevant to hydrogen-rich amorphous/crystalline interfaces in SHJ cells. Beyond SHJ cells, hydrogen-related defect processes may also be relevant in other photovoltaic materials. In metal-halide perovskites, hydrogen interstitials have been predicted to contribute to nonradiative recombination, while the formation of electrically inactive \(\ce{H2}\) might reduce these losses \cite{Liang2022HydrogenPerovskite}. Hydrogen interstitials have also been predicted to migrate to and accumulate at \(\mathrm{FAPbI_3}\) surfaces, where they can increase nonradiative recombination. The double-H mechanism identified here may also be relevant to the hydrogen-passivated silicon subcell in perovskite/silicon tandem devices. More broadly, the workflow developed here can be applied to other photovoltaic materials.

% ============================================================
\section*{Acknowledgments}
% ============================================================
The authors acknowledge funding from the Helmholtz-Gemeinschaft Deutscher Forschungszentren e.V. (HGF), Program-oriented Funding (PoF IV), under the Research Program: Materials and Technologies for the Energy Transition (MTET). The authors gratefully acknowledge the Gauss Centre for Supercomputing e.V. (www.gauss-centre.eu) for funding this project by providing computing time through the John von Neumann Institute for Computing (NIC) on the GCS Supercomputer JUWELS at Jülich Supercomputing Centre (JSC). H.A. acknowledges support from the HITEC Fellowship 2024. The authors also thank Andrew Diggs (University of California, Davis (UC Davis)), Alexander Eberst (IMD-3, Jülich), Andreas Lambertz (IMD-3, Jülich), Weiyuan Duan (IMD-3, Jülich), Dustin Vivod (IET-3, Jülich) and Michael Dürr (Justus Liebig University Giessen) for helpful discussions.

\pagebreak

\clearpage
\onecolumngrid
\appendix

\section{Hybrid Functional Projected Density of States}
\label{app:pdos}

\begin{figure}[H]
    \centering
    \includegraphics[width=\textwidth]{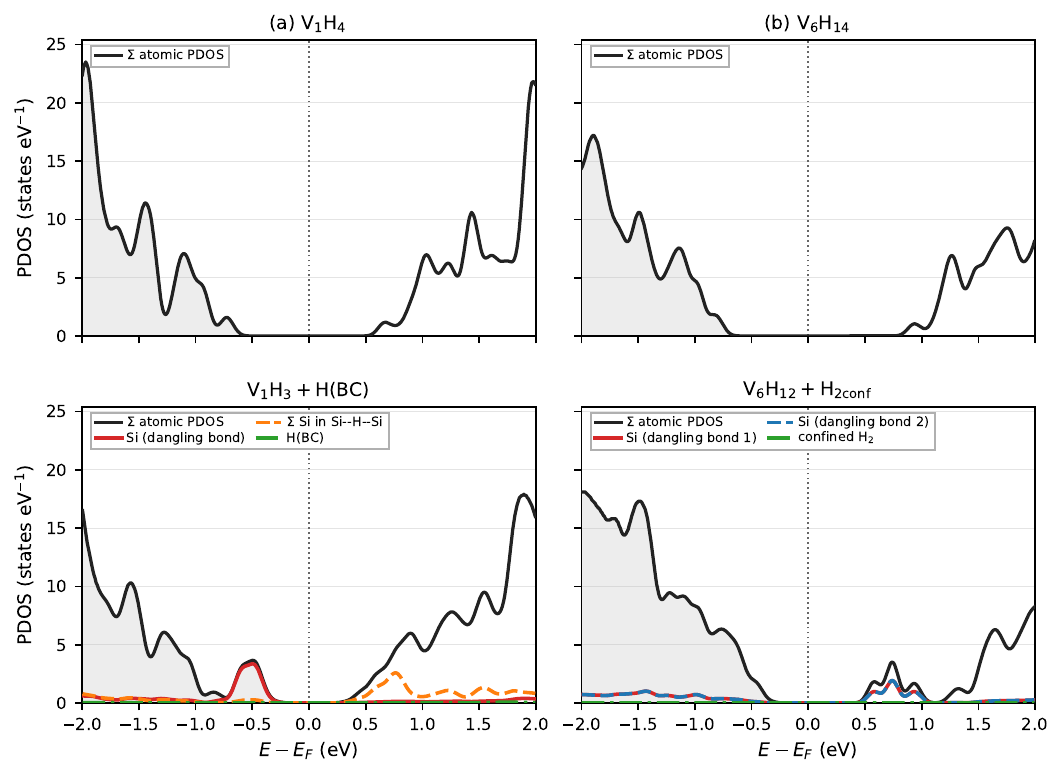}
    \caption{HSE06 projected densities of states for the (a) single-H and (b) double-H depassivation processes. The upper panels show the fully passivated initial states, and the lower panels show the corresponding depassivated final states. The vertical dashed line denotes the Fermi level \(\mathrm{E_F}\).}
    \label{fig:pdos_hse}
\end{figure}

The HSE06 projected densities of states for the passivated and depassivated configurations are shown in Fig.~\ref{fig:pdos_hse}. The passivated configuration shows a clean band gap, whereas the depassivated single-H and double-H final states contain occupied or empty states within the band gap.

Figure~\ref{fig:pdos_hse} provides a hybrid functional check of the PBE results. HSE06 increases the separation between the occupied and unoccupied states near \(\mathrm{E_F}\), widening the residual gap from $0.204$ to $0.637\,\mathrm{eV}$ for $\mathrm{V_1H_3+H(BC)}$ and from $0.380$
to $1.086\,\mathrm{eV}$ for $\mathrm{V_6H_{12}+H_{2\mathrm{conf}}}$. This widening is expected due to the screened exact exchange included in HSE06, which partially corrects the self-interaction error present in semilocal PBE and thereby improves the description of the band gap in semiconductors. HSE06 also preserves the defect states found with PBE, with the dangling-bond character in the fully occupied HOMO for single-H and the empty LUMO for double-H.

\newpage
\section{Single-H depassivation of the \(\mathrm{V_6}\) cavity}
\label{app:v6_single_h}

\begin{figure}[H]
\centering

\begin{minipage}[t]{0.32\textwidth}
    \textbf{(a)}\\
    \includegraphics[width=\linewidth]{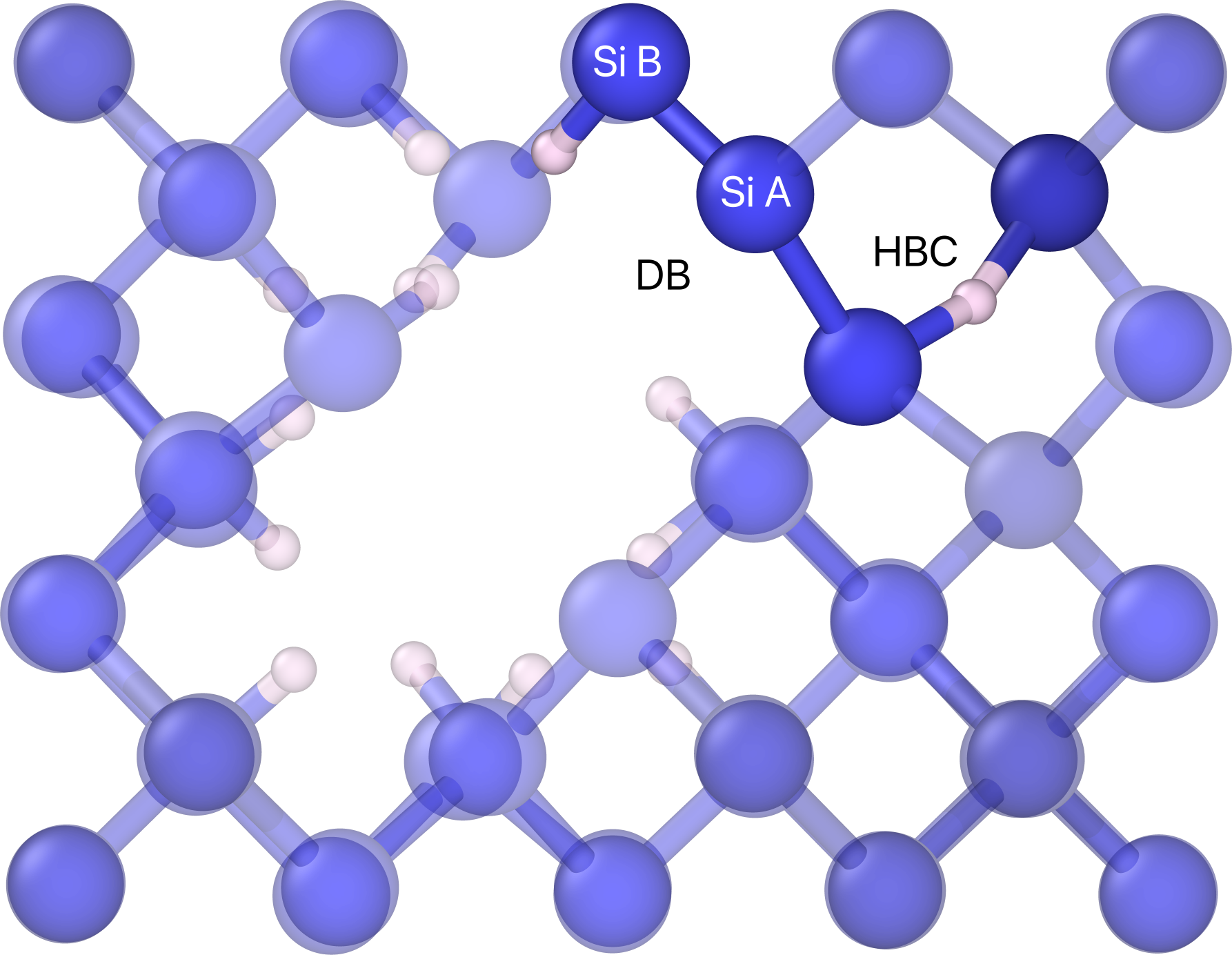}\\
    \small p-type
\end{minipage}
\hfill    
\begin{minipage}[t]{0.32\textwidth}
    \textbf{(b)}\\
    \includegraphics[width=\linewidth]{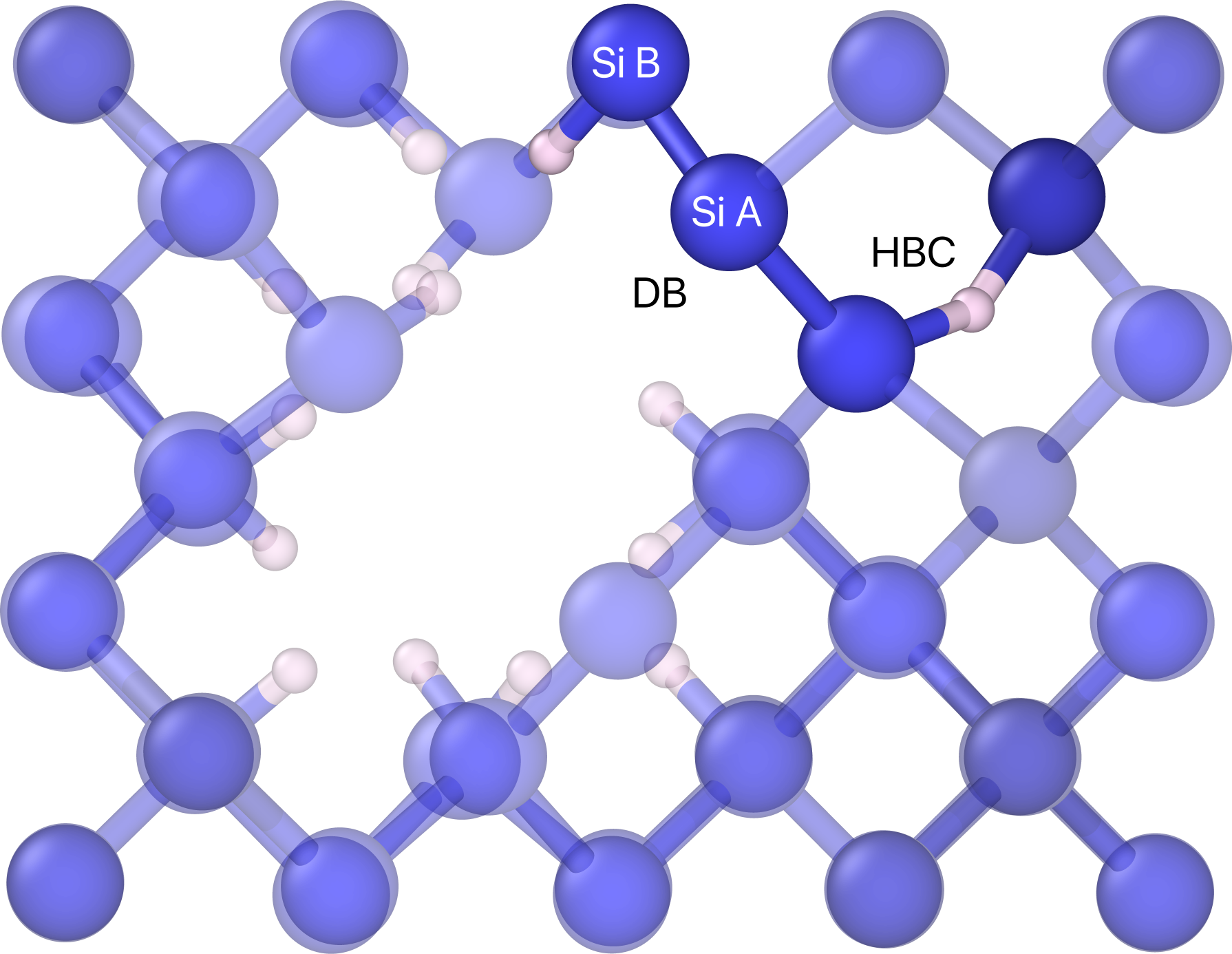}\\
    \small undoped
\end{minipage}
\hfill 
\begin{minipage}[t]{0.32\textwidth}
    \textbf{(c)}\\
    \includegraphics[width=\linewidth]{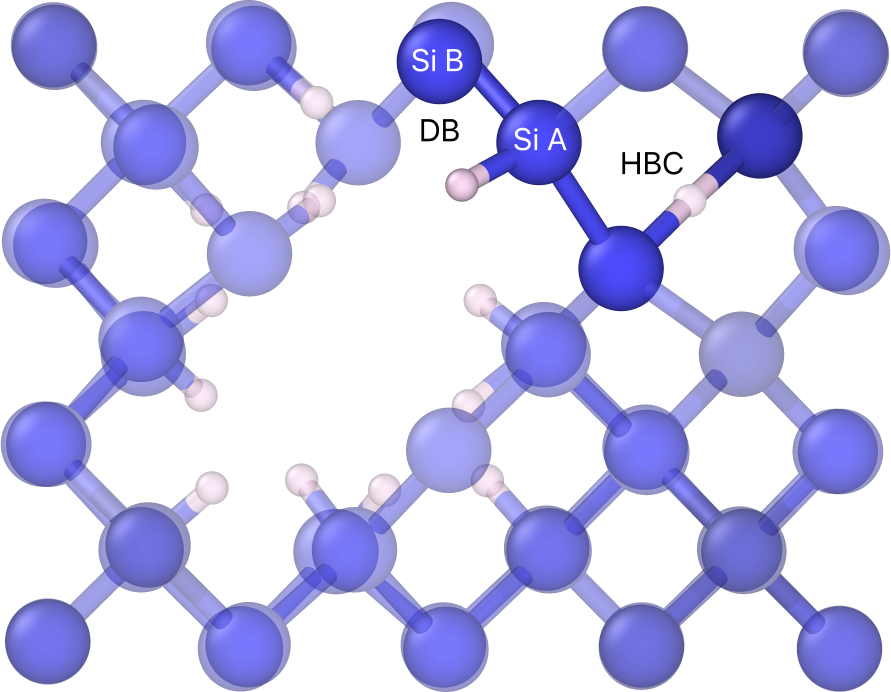}\\
    \small n-type
\end{minipage}

\caption{\label{fig:v6_single_h}
Relaxed single-H depassivated \(\mathrm{V_6}\) configurations under (a) p-type, (b) undoped, and (c) n-type conditions. A cavity-wall H atom passivates Si atom A in the p-type and undoped structures, but Si atom B in the n-type structure.}
\end{figure}

For comparison with the double-H pathway, we also considered single-H depassivation of the \(\mathrm{V_6}\) cavity,
\begin{equation}
    \mathrm{V_6H_{14}}
    \rightarrow
    \mathrm{V_6H_{13}}+\mathrm{H(BC)}.
\label{eq:v6_single_h}
\end{equation}
In this process, one Si--H bond on the cavity wall is broken, creating one dangling bond, while the released H atom occupies a bond-centered interstitial site. The single-H depassivation reaction energy is \(1.955\)~eV, which is \(0.381\)~eV higher than the corresponding double-H depassivation energy of \(1.574\)~eV. Thus, despite involving the breaking of two Si--H bonds, the double-H depassivation process has a lower reaction energy than the single-H process in the \(\mathrm{V_6}\) cavity. The single-H final state was also relaxed under n-type and p-type conditions to examine the doping effect on the local structure. The relaxed structures are shown in Fig.~\ref{fig:v6_single_h}. In all three cases the released H atom remains coordinated to two Si atoms in a Si--H--Si bridge. The H atom is close to the bond center between the two Si atoms in the n-type and p-type structures, whereas the undoped structure shows a larger bond-length asymmetry.

The relaxed structures also show a local rearrangement at the cavity wall in the n-type case. As shown in Fig.~\ref{fig:v6_single_h}, the H atom that passivated Si atom A in the undoped and p-type structures instead passivates Si atom B in the n-type structure. Consequently, the dangling bond site depends on the doping condition. Since the single-H process is also higher in energy than the double-H process and the relaxed final structures are inconsistent under different doping conditions, the single-H \(\mathrm{V_6}\) configurations are included only as a secondary comparison. The main analysis therefore focuses on the double-H process involving confined \(\mathrm{H_2}\).

\FloatBarrier
\clearpage
\twocolumngrid

% ============================================================
\bibliography{references}
% ============================================================

\end{document}